\documentclass[twocolumn]{aastex631}

\usepackage{amsmath}
\usepackage{CJKutf8}
\usepackage{enumitem}

\shorttitle{Feedback-Driven Star Formation in the Circinus Complex}
\shortauthors{Kerr et al.}

\begin{document}

\title{SPYGLASS. VI. Feedback-Driven Star Formation in the Circinus Complex}

\defcitealias{Kerr21}{SPYGLASS-I}
\defcitealias{Kerr22b}{SPYGLASS-II}
\defcitealias{Kerr22a}{SPYGLASS-III}
\defcitealias{Kerr23}{SPYGLASS-IV}
\defcitealias{Kerr24}{SPYGLASS-V}

\correspondingauthor{Ronan Kerr}
\email{ronan.kerr@utoronto.ca}

\author[0000-0002-6549-9792]{Ronan Kerr}
\affiliation{Dunlap Institute for Astronomy \& Astrophysics, University of Toronto,
Toronto, ON M5S 3H4, Canada\\}

\author[0000-0002-5851-2602]{Juan P. Farias}
\affiliation{Department of Astronomy, University of Texas at Austin,
2515 Speedway, Stop C1400, Austin, Texas, USA 78712-1205\\}

\author[0000-0001-7998-226X]{Lisa Prato}
\affiliation{Lowell Observatory,
1400 West Mars Hill Road, Flagstaff, AZ 86001 USA}

\author[0000-0001-8164-653X]{Travis A. Rector}
\affiliation{University of Alaska Anchorage, Department of Physics \& Astronomy,
Anchorage, AK 99508, USA}

\author[0000-0003-2573-9832]{Joshua S. Speagle (\begin{CJK*}{UTF8}{gbsn}沈佳士\ignorespacesafterend\end{CJK*})}
\affiliation{Dunlap Institute for Astronomy \& Astrophysics, University of Toronto,
Toronto, ON M5S 3H4, Canada\\}
\affiliation{Department of Statistical Sciences, University of Toronto\\ 9th Floor, Ontario Power Building, 700 University Ave, Toronto, ON M5G 1Z5, Canada}
\affiliation{David A. Dunlap Department of Astronomy \& Astrophysics, University of Toronto\\ 50 St George Street, Toronto, ON M5S 3H4, Canada}
\affiliation{Data Sciences Institute, University of Toronto\\ 17th Floor, Ontario Power Building, 700 University Ave, Toronto, ON M5G 1Z5, Canada} 

\author[0000-0001-9811-568X]{Adam L. Kraus}
\affiliation{Department of Astronomy, University of Texas at Austin, 
2515 Speedway, Stop C1400, Austin, Texas, USA 78712-1205\\}



\begin{abstract}

Young associations provide a record that traces the star formation process, and the youngest populations connect progenitor gas dynamics to the resulting stellar populations. We therefore conduct the first comprehensive overview of the Circinus Complex, an under-studied and massive ($\sim$1500 M$_{\odot}$) region consisting of approximately 3100 recently formed stars alongside the Circinus Molecular Cloud (CMC).  We find a clear age pattern in the contiguous central region (CirCe), where younger stars are found further from the massive central cluster, and where the velocities are consistent with uniform expansion. By comparing this structure to an analogous STARFORGE simulation, we find that the age structure and dynamics of the association are consistent with star formation in two stages: the global collapse of the parent cloud that builds the $500 M_{\odot}$ central cluster ASCC 79, followed by triggered star formation in a shell swept up after the first massive stars form. We also find that filaments with a range of distances from the central cluster can naturally produce multi-generational age sequences due to differences in feedback strength and exposure. Outlying populations show velocities consistent with formation independent from the CirCe region, but with similar enough velocities that they may be difficult to distinguish from one another later in their expansion. We therefore provide a new alternative view of sequential star formation that relies on feedback from a single central cluster rather than the multiple sequential generations that are traditionally invoked, while also providing insight into the star formation history of older populations. 

\end{abstract}


\section{Introduction} \label{sec:intro}

Most nearby young stars reside in largely unbound stellar associations that record the history of the star forming event that produced them \cite[e.g.,][]{deZeeuw99, Krumholz19}. By combining the dynamics and ages of the stars in these associations, we can reconstruct the times and locations where stars formed throughout the evolution of the parent cloud, revealing patterns such as age gradients \citep[e.g.,][]{Pang21, Ratzenbock23, Kerr24}, discrete star formation nodes \citep[e.g.,][]{Kerr22a, Kerr22b}, and acceleration signatures \citep[e.g.,][]{Posch24}. 

One topic that has gained increased attention lately is the concept of sequential star formation, which has evolved into a term to describe a range of mechanisms that lead to the formation of age gradients in nearby associations. Sequential star formation was first proposed by \citet{Elmegreen77}, who suggested that age gradients can be produced through a repeating cycle of star formation, compression of adjacent material through feedback, and initiation of more star formation in the adjacent compressed material \citep{Shu87}. \citet{Posch23} recently proposed a new form of sequential star formation in which a cloud adjacent to a ``progenitor cluster'' is continuously accelerated by stellar feedback and supernovae. This produces a chain of clusters with progressively higher velocities and younger ages further away from the progenitor cluster, driven by the combined feedback of the progenitor cluster and any other generations that precede the current generation. However, the limited sample size of populations that host these patterns, combined with the lack of corroborating simulations, limit our ability to confidently tie patterns in young associations to specific guiding processes. 

The widespread availability of stellar proper motions and distances though \textit{Gaia} has improved our ability to detect young stars and identify co-moving populations within. Several surveys of nearby stellar populations have been published using this new dataset, including many surveys that reveal star clusters of all ages \citep{Kounkel19, Sim19, Hunt23}, and some that focus on young clusters in particular, taking advantage of the photometric differences between young and old stars to minimize field contamination during clustering \citep{Prisinzano22}. The SPYGLASS (Stars with Photometrically Young Gaia Luminosities Around the Solar System) Program is one such paper series focused on young stellar populations \citep{Kerr21}. The most recent SPYGLASS survey paper, \citet{Kerr23} (hereafter SPYGLASS-IV), revealed 116 young associations within 1 kpc, including 30 with no direct literature equivalents, and several highly substructured stellar complexes with minimal literature coverage. SPYGLASS regional follow-up studies such as \citet{Kerr22b} (SPYGLASS-II) in Cepheus Far North, \citet{Kerr22a} (SPYGLASS-III) in the Austral Complex, and \citet{Kerr24} (SPYGLASS-V) in the Cepheus-Hercules complex have revealed patterns of multi-generational, multi-origin star formation, as well as age gradients that may contribute to our understanding of sequential star formation. Studies of often-overlooked young populations help to complete our view of the processes guiding local star formation, and the effects those processes may have on the stars or planets that emerge. 

The Circinus Complex is one such under-explored region. It has typically been recognized by its connection to the Circinus Molecular Cloud (CMC), which consists of two adjacent and potentially contiguous clouds, referred to as Circinus East and West (hereafter, Cir-E and Cir-W) \citep{Reipurth08-cir}. These regions host abundant dense molecular gas and show evidence for active star formation, which is supported by the presence of protostars and Herbig-Haro objects, especially in Cir-W \citep{Reipurth08-cir, Rector20}. An additional molecular gas cloud hosting potential dense cores was identified to the galactic north of these protostar-hosting regions, although this cloud has seen no dedicated research, nor has it been assigned a name \citep{Dobashi05, Reipurth08-cir}. The dynamics and evolutionary state of the region are therefore unclear. Several stellar populations are found adjacent to these gas structures, with the largest being ASCC 79 \citep{Kharchenko05}, which is found alongside several other smaller clusters such as UPK 604, 607, and 610 \citep{Sim19}. These clusters and the CMC all lie at a similar distance, which most publications place between 700 and 900 pc \citep[e.g.,][]{Kharchenko12, CantatGaudin20, Zucker20, Kerr23}. The Circinus OB1 association is found along essentially the same sight line, although it appears to be much further away, at about 3 kpc \citep{VanderHucht01}. No publication to date has explored the connections between these clusters and the adjacent gas content, nor has any dedicated research investigated the complete extent and structure of the young populations in the region.

The Circinus Complex was first cataloged as a single, unified population in \citetalias{Kerr23}, where a large region that includes the CMC was grouped together under the population ID SCYA-6. This view of the Circinus region shows a clear pre-main sequence, although it also shows strong evidence for field star contamination that may be overwhelming and obscuring substructure. In this publication, we revisit the \citetalias{Kerr23} sample for Circinus, reassess its membership to reduce contamination, and provide the first comprehensive study of the structure and star formation history of the region. In Section \ref{sec:data} we discuss the dataset used for this analysis. We then discuss Circinus' substructure in Section \ref{sec:structure}, stellar masses and OB stars in Section \ref{sec:obstars}, ages in Section \ref{sec:ages}, demographics in Section \ref{sec:demographics}, and dynamics in Section \ref{sec:dynamics}. We compare the properties of the Circinus complex with an analogous STARFORGE simulation in Section \ref{sec:starforge}, producing a plausible explanation for the formation sequence of the association. We provide some supplemental discussion in Section \ref{sec:discussion}, before concluding in Section \ref{sec:conclusion}. 

\section{Data} \label{sec:data}

\subsection{Gaia Candidate Member Selection} \label{sec:basesample}

\begin{figure}
\centering
\includegraphics[width=8cm]{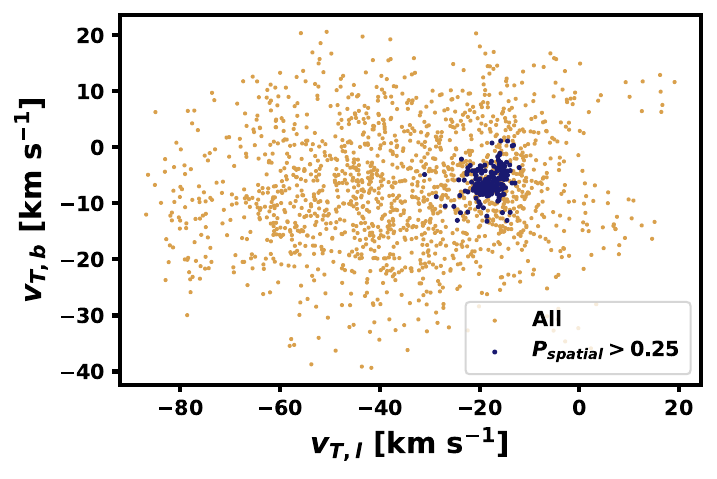}\hfill
\caption{Transverse velocities (as described in Sec. \ref{sec:basesample}) for all photometrically young members of Circinus listed in \citetalias{Kerr23} (tan), which is dominated by heavily-reddened subgiants, plotted against photometrically young stars that pass our $P_{spatial}$ cut (dark blue), which excludes most field contaminants.}
\label{fig:circinusvel}
\end{figure}

Our list of Circinus candidate members was drawn from \citetalias{Kerr23}. Using a Bayesian framework for computing stellar youth probability based on PARSEC v1.2S stellar evolution models \citep{PARSECChen15}, \citetalias{Kerr23} identified a sample of over $4\times10^5$ photometrically young ($P_{Age<50 Myr}>0.2$) stars, which were clustered into 116 young stellar populations using the HDBSCAN clustering algorithm \citep{McInnes2017}. The Circinus Complex was given the catalog ID SCYA-6, and consists of 1756 photometrically young stars used to define the population, along with 261606 total candidate members, which includes space-velocity neighbors that share the positions and dynamics of the photometrically young members but do not have a conclusive photometric membership assessment. These values almost certainly overestimate the size of the population, however, due to the presence of extinction-induced field contamination. \citetalias{Kerr23} uses the \citet{Lallement19} maps to correct for extinction, however the CMC is sufficiently dense and substructured that these maps cannot account for the small scale, high-density gas clouds found in the region. This causes heavily-extincted field subgiants to get pushed onto the pre-main sequence and marked as photometrically young members. As a result, the transverse velocities, or the sky-plane velocities implied by the star's proper motion and distance, shows a distribution that combines both Circinus and the field, as shown in  Fig. \ref{fig:circinusvel}. The true association members can be found primarily in the overdensity with transverse velocities centered on $(v_{T,l},v_{T,b}) \sim (-18, -7)$. This corresponds to a median proper motion of $(\mu_{RA}, \mu_{Dec}) = (-3.1, -4.0)$ mas yr$^{-1}$, which is similar to the value of $(\mu_{RA}, \mu_{Dec}) = (-2.9, -4.2)$ mas yr$^{-1}$ reported for ASCC 79, the main clustered population in the region \citep{CantatGaudin20}. 

The extensive field contamination in Circinus necessitates restrictive cuts to make the dataset operable. We therefore impose a cut on the spatial membership probability, $P_{spatial}$ (referred to as $P_{mem}$ in \citetalias{Kerr23}). This metric compares the relative frequencies of photometrically young and old stars, and measures the fraction of stars at a given location in space-transverse velocity coordinates that are likely to be members of the parent association. We require $P_{spatial} > 0.25$ for our analysis, which removes most of the field velocity component and limits the population to the dense clump around $(v_{T,l},v_{T,b}) = (-18, -7)$ that we associate with the Circinus Complex. The restricted selection is overlaid on the full transverse velocity distribution in Figure \ref{fig:circinusvel}. This restriction reduces the original 261606-star candidate sample to a much smaller set of 3471 stars that we use for the rest of this paper. We list all candidate members in Table \ref{tab:members}. 

The cut on $P_{spatial}$ uses \textit{Gaia} positions and velocities to substantially limit contamination in the candidate member sample. However, while this sample is useful for demographic purposes, not all of its stars are photometric members, and some of them do not have properties that support analysis. We must therefore produce a restricted list of high-quality members that can be used to reveal substructure and compute ages. 

\begin{deluxetable*}{cccccccccccccc}
\tablecolumns{14}
\tablewidth{0pt}
\tabletypesize{\scriptsize}
\tablecaption{List of Circinus Complex members, including IDs, basic properties, substructure assignments, and membership probabilities. The complete version of this table is available in the online-only version, which contains 3471 members.}
\label{tab:members}
\tablehead{
\colhead{Gaia DR3 ID} &
\colhead{CIR\tablenotemark{a}} &
\colhead{RA} &
\colhead{Dec} &
\colhead{d\tablenotemark{b}} &     
\colhead{g} &  
\colhead{M} &  
\colhead{F\tablenotemark{c}} &  
\colhead{$P_{\rm Age<50Myr}$} &  
\colhead{$P_{\rm spatial}$} &
\colhead{$P_{\rm sp}$} &
\multicolumn{3}{c}{RV} \\
\colhead{} &
\colhead{} &
\colhead{(deg)} &
\colhead{(deg)} &
\colhead{(pc)} &
\colhead{} &
\colhead{(M$_{\odot}$)} &
\colhead{} &
\colhead{} &
\colhead{} &
\colhead{} &
\colhead{val} &
\colhead{err} &
\colhead{src}\\
}
\startdata
5825256004656218112 & 4 & 228.4869 & -64.4397 & 762 & 19.26 & 0.49 & 2 &  & 0.33 &  &  &  &  \\
5825260711928978944 & 4 & 228.9856 & -64.3488 & 760 & 20.9 & 0.27 & 0 &  & 0.28 &  &  &  &  \\
5825263353300485120 & 4 & 228.9325 & -64.3395 & 772 & 19.67 & 0.45 & 2 &  & 0.27 &  &  &  &  \\
5825300496174143744 & 4 & 230.1319 & -64.2159 & 799 & 20.15 & 0.4 & 2 &  & 0.27 &  &  &  &  \\
5825346508209042176 & 4 & 229.8145 & -64.2413 & 806 & 15.01 & 0.94 & 0 & 0.0021 & 0.27 & 0.14 & -16.76 & 9.5 & GDR3 \\
5825350493908583424 & 4 & 229.5994 & -64.1705 & 806 & 20.71 & 0.36 & 0 &  & 0.28 &  &  &  &  \\
5825352246254160768 & 4 & 229.7295 & -64.1378 & 778 & 19.83 & 0.43 & 0 &  & 0.25 &  &  &  &  \\
5825368910752816384 & 4 & 229.5223 & -63.9716 & 793 & 16.1 & 0.76 & 0 & 0.0009 & 0.39 & 0.12 &  &  &  \\
5825369391770726016 & 4 & 229.5389 & -63.9217 & 779 & 20.77 & 0.35 & 0 &  & 0.27 &  &  &  &  \\
5825369765402134656 & 4 & 229.5048 & -63.8938 & 791 & 12.12 & 1.74 & 0 & 0.0081 & 0.4 & 0.54 & -32.76 & 19.57 & GDR3 \\
 \enddata
\vspace*{0.1in}
\tablenotetext{a}{Circinus parent subgroup ID} 
\tablenotetext{b}{\citet{BailerJones21} geometric distance}
\tablenotetext{c}{Binarity flag: 1 indicates that the star has a resolved companion within 10,000 au in the plane of the sky, 2 indicates RUWE$>$1.2, indicating likely unresolved binarity. The flags are added in cases where both are true.} 
\end{deluxetable*}

To introduce photometry into our membership assessment, we compute the spatial-photometric membership probability, $P_{sp}$, as defined in \citetalias{Kerr24}. This metric combines the photometric youth probability $P_{Age < 50 Myr}$, which is calculated in \citetalias{Kerr23} assuming density of young stars consistent with the field, with $P_{spatial}$, which provides the fractions of young and old stars near a given point in space-velocity coordinates. The spatial-photometric probability, $P_{sp}$, therefore adjusts the youth fraction prior on $P_{Age < 50 Myr}$ to align with the youth fraction indicated by $P_{spatial}$. Equations 1-3 in \citetalias{Kerr24} provide the formulae used for this calculation. $P_{sp}$ serves as our final assessment of membership probability, and we therefore use it to restrict our high-quality member sample, requiring $P_{sp}>0.5$.

We also apply cuts to remove stars with unreliable \textit{Gaia} measurements. First, we require that each star passes the photometric and astrometric quality flags defined in \citetalias{Kerr23}. These restrictions are based on those recommended in \citet{Arenou18}, which use the astrometric $\chi^2$ and number of good visits to assess \textit{Gaia} astrometric quality, and use the $BP-RP$ excess factor to flag inconsistent photometric solutions. We also restrict on the \citet{BailerJones21} geometric distance uncertainty, requiring $\sigma_d < 100$ pc. Parallax measurements are the largest source of uncertainty in stellar positions, so removing stars with particularly uncertain distances makes it easier to detect structures in 3D space.

\begin{figure}
\centering
\includegraphics[width=8cm]{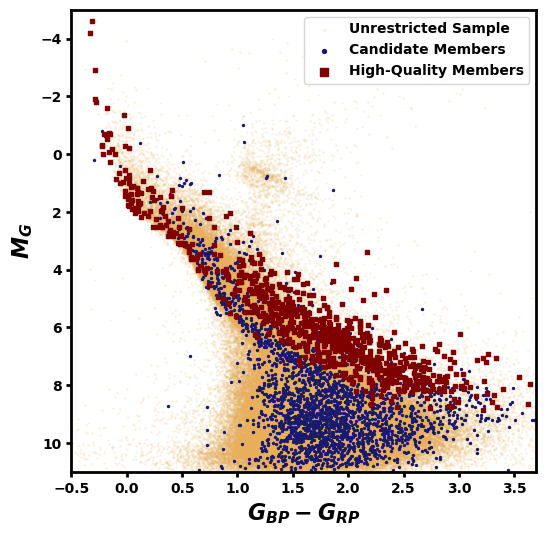}\hfill
\caption{Color magnitude diagram containing the complete \citetalias{Kerr23} sample for Circinus (tan), the $P_{spatial}$-restricted sample (dark blue), and the high-quality member sample (dark red). The high-quality member sample contains only photometrically young stars with positions and velocities consistent with the Circinus Complex, and these stars form a clear pre-main sequence elevated above most of the stars in the other two samples.}
\label{fig:circinuscmd}
\end{figure}

Stars that survive the restrictions to $P_{sp}$, the astrometric and photometric quality flags, and the distance uncertainty define our high-quality member sample. We present a color-magnitude diagram showing the distribution of high-quality members in Figure \ref{fig:circinuscmd}, alongside the original unrestricted \citetalias{Kerr23} sample and the $P_{spatial}$-restricted set of candidate members. We show that the candidate member population is consistent with a purer subset of the unrestricted population, while the high-quality member sample contains most stars that would fall near a typical pre-main sequence. The high-quality sample only systematically excludes credible members near the bottom of the pre-main sequence, where stars routinely fail our quality cuts, and near the pre-main sequence turn-on, where young and old stars are photometrically near-identical. Some stars that fall near the field main sequence survive this cut, which is expected behavior for stars near the center of the region's space-velocity distribution that have sufficiently ambiguous photometric youth. 

\begin{figure*}
\centering
\includegraphics[width=17cm]{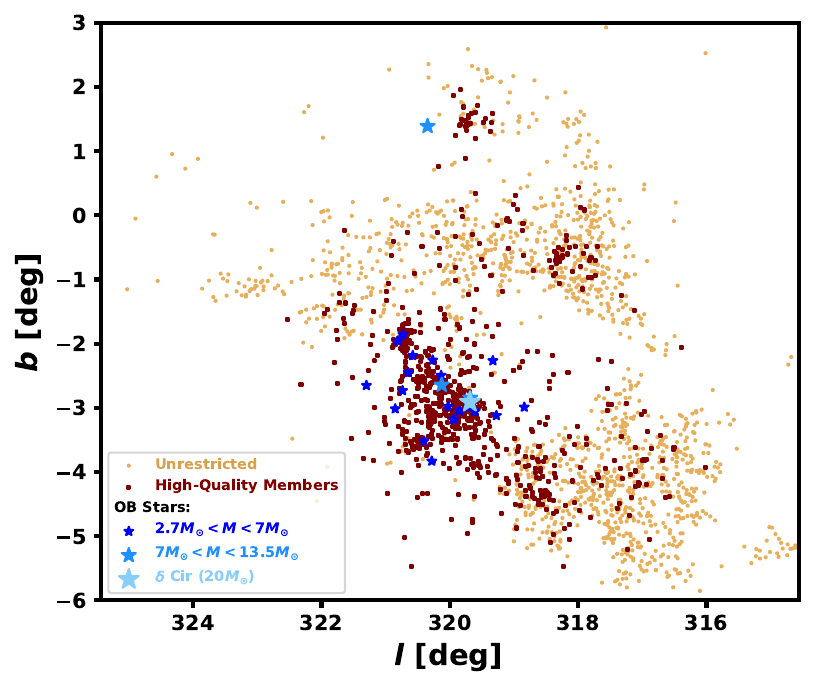}\hfill
\caption{Sky distribution of our high-quality member sample used for structural analysis (dark red, matching Fig. \ref{fig:circinuscmd}), plotted against the photometrically young heavily-reddened stars removed in Section \ref{sec:data}, which have velocities broadly consistent with field contaminants (tan). The distribution of these contaminants is consistent with the known distribution of dust in the area. We also mark O and B stars with blue star markers. The size and shade corresponds to their mass calculated by comparison to PARSEC isochrones, ranging from small dark markers for stars with $M < 7 M_{\odot}$, to large light markers for the $\delta$ Cir system, which has a $20 M_{\odot}$ primary.}
\label{fig:circinussky}
\end{figure*}

The distribution of the high-quality members on-sky is presented in Figure \ref{fig:circinussky}, revealing that most high-quality members reside in a large substructured complex centered on $(l, b) = (320, -3)$, with a few smaller populations distributed beyond that central clump. We plot these members against the sample of photometrically young candidates from \citetalias{Kerr23} prior to the $P_{spatial}$ restriction, which are dominated by heavily-reddened subgiants \citep{Kerr23}. As such, the distribution of these candidates is a near-direct tracer of molecular gas, their locations matching closely with the dust map from \citet{Dobashi05}. Parts of these heavily reddened regions appear to be curved around the center of Circinus, indicating that the adjacent gas structures may be actively sculpted by the young stellar populations found there. 

\subsection{Supplemental Data} \label{sec:supdat}

Spectroscopy is useful for studies of young stellar populations, as it provides youth indicators like the Li 6708 \AA~line and the H$\alpha$ emission line, in addition to radial velocities, which facilitate determination of the 3-D velocity vector of association members and make it possible to dynamically distinguish between members and non-members. We therefore perform a literature search for spectroscopy in the Circinus complex. We search the same set of sources queried in \citetalias{Kerr24}, in addition to the Gaia-ESO Survey and APOGEE DR19, which was accessed as SDSS-V IDL-3. This cross-match revealed that none of the 3471 Circinus candidate members in our restricted sample have measurements of any spectroscopic youth indicators. Three stars in the unrestricted set of 261606 candidates have Lithium measurements, and none of them have $EW_{Li}>0.1$ \AA, so our initial downselect did not remove any spectroscopically confirmed young stars.

Radial velocities were available for 329 stars, with 327 of these being from Gaia DR3, and the remaining two coming from \citet{Gontcharov06}. However, despite this coverage, nearly all of the radial velocities in the area have high uncertainties. Only 11 stars in the sample have sub-km s$^{-1}$ uncertainties, and the average uncertainty across all candidates is 7.6 km s$^{-1}$. At the distance of Circinus, this is not surprising, as the bright early-type stars with \textit{Gaia} measurements are typically fast rotators at this age \citep{Weise10}, and they often host accretion and activity features that are not considered in the \textit{Gaia} RV fits, and contribute to measurement scatter that can exceed \textit{Gaia} uncertainties \citep{Kounkel22}. Average RV uncertainties increase with the purity of the young sample, with mean uncertainties of 5.83 km s$^{-1}$ among the original 261606-member sample, and 8.25 km s$^{-1}$ among high-quality members, indicating that stars with low uncertainties have a disproportionately high chance of being field contaminants. The available RV data therefore cannot reliably assess the membership of individual stars, although they do have utility for computing bulk velocity vectors at the scale of the entire population, which can help quantify the contribution of geometric effects to the cluster expansion signature \citep{Brown97}. We therefore add the RVs to our sample data.

\section{Substructure} \label{sec:structure}

To deepen our understanding of Circinus' structure, we cluster the high-quality member sample using the HDBSCAN clustering algorithm \citep{McInnes2017}, which uses a minimum spanning tree to identify stellar structures with minimal input bias for population size and morphology. However, unlike previous associations covered by SPYGLASS, the Circinus association is quite distant ($\sim 900$ pc in \citetalias{Kerr23}), resulting in uncertainties that stretch the population substantially in the radial dimension. Furthermore, since the population's diameter is only $\sim 5^{\circ}$ on-sky, the geometric effects from one side of the association to the other are limited. Therefore, rather than clustering in 5D XYZ galactic cartesian coordinates and transverse velocity, we cluster in galactic sky coordinates, distance, and transverse velocity, and readjust the scales of each to align with the elevated dispersions produced by the large uncertainties, as described below. 

\begin{figure}
\centering
\includegraphics[width=8cm]{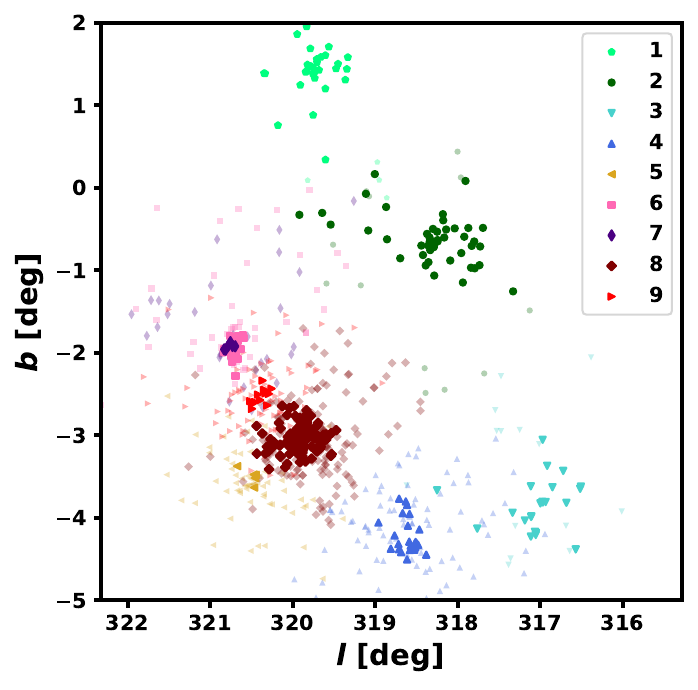}\hfill
\caption{Assignment of outlying members in Circinus, shown in the $l$ vs $b$ sky plane. Stars assigned to a group by HDBSCAN clustering are shown in the dark shades listed in the legend, while stars initially assigned to the background component are presented in faded shades, with the shape and color matching the group we later assign them to.}
\label{fig:outlyingassignment}
\end{figure}

\begin{figure*}
\centering
\includegraphics[width=18cm]{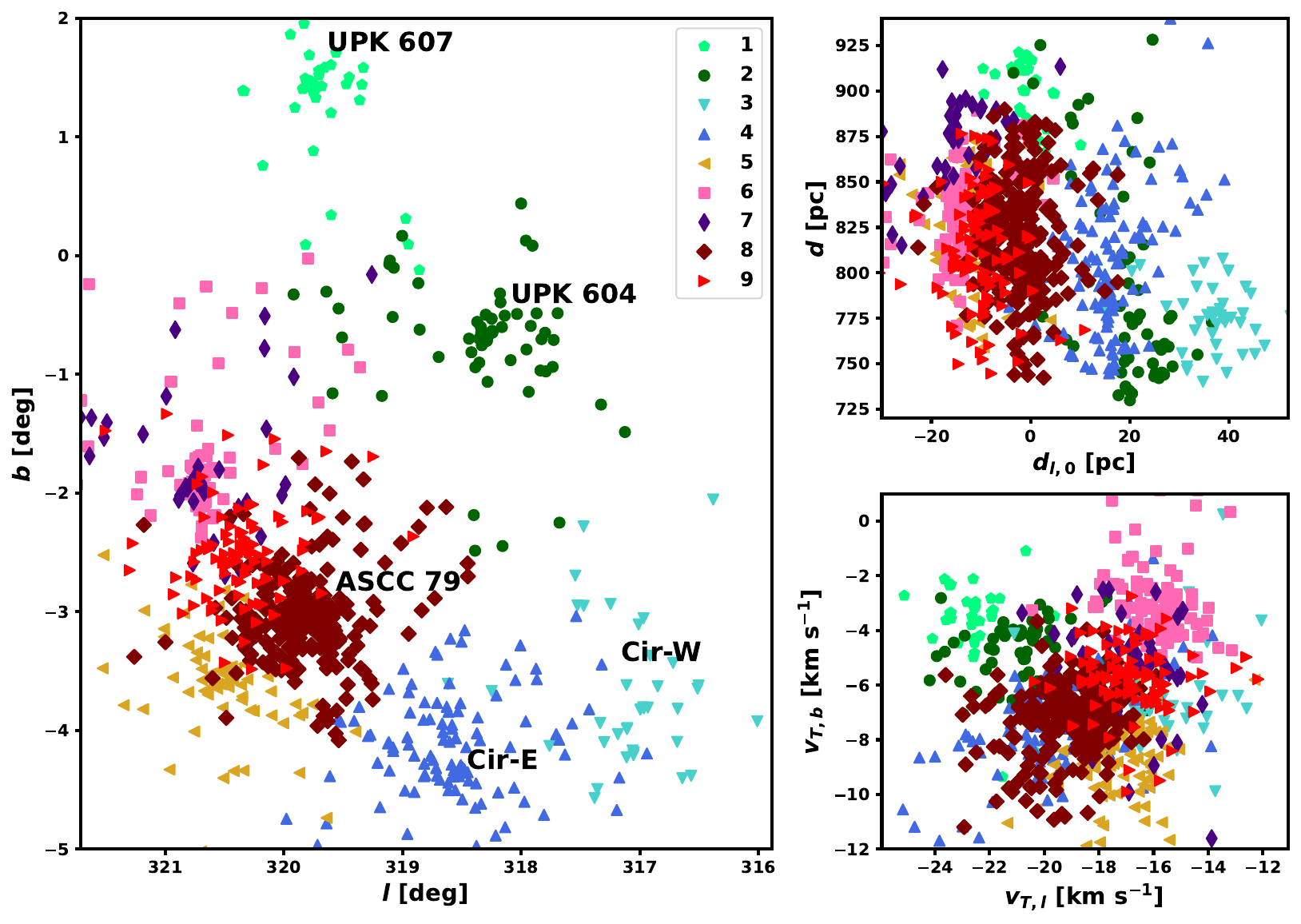}\hfill
\caption{Subgroups identified with HDBSCAN in Circinus. In the left panel, we plot these populations in $l$ vs $b$ galactic sky coordinates, while we show the distribution of stars in distance versus sky-plane distance in the $l$ direction relative to the mean in the top-right panel, and the transverse velocity distribution in the bottom-right panel.}
\label{fig:circinus}
\end{figure*}

The clump centered near $(l,b) = (319.7\degr, 1.5\degr)$ has a similar on-sky extent to other visibly identifiable groups, and can be easily isolated with the cut of $b>0.5\degr$, making it a plausible reference population for informing this scaling. This clump has a distance standard deviation of 17.8 pc, compared to 3.5 pc in the $l$ direction and 3.4 pc in the $b$ direction. It also has standard deviations of 0.85 km s$^{-1}$ and 0.75 km s$^{-1}$ in $l$ and $b$ transverse velocity, respectively. The choice whether to include spatial outliers in these statistics changes the velocity and distance spread by no more than $\sim$10\%, however the spatial extent is more sensitive, varying by about a factor of 2. To determine whether this population and its scale are representative of subgroups in associations like Circinus, we compare it to Upper Sco, the nearest stellar population with a similar size and age. This population has subgroup 1-$\sigma$ radial extents that average 3.7 pc once the on-sky extents from \citetalias{Kerr21} are converted from sky coordinates to physical extent, very similar to the extent measured for our reference clump. The 1D velocity dispersions for Upper Sco in \citetalias{Kerr21} average 0.44 km s$^{-1}$ in the major axis and 0.34 km s$^{-1}$ in the minor axis. However, our test clump has much larger transverse velocity uncertainties, which average 0.76 km s$^{-1}$. The expected velocity dispersion of a population at the location of this clump is provided by the addition in quadrature of its intrinsic velocity dispersion and its measurement uncertainty, which would be 0.83 km s$^{-1}$ on the minor axis and 0.87 km s$^{-1}$ on the major axis for Upper Sco, aligning closely with the velocity dispersions for the clump. The 17.8 pc distance spread in our test clump is much larger than the typical size of an Upper Sco population, however due to the distance, this scale is dominated by the parallax uncertainties, not the true subgroup size. We therefore conclude that this clump has properties typical of subgroups in populations like Circinus, and we use its properties to set the scaling factors used for clustering.

To scale the sky coordinates to match distance, we first convert from sky coordinates to linear distance in the plane of the sky ($d_{l,0}$, $d_{b,0}$), measured relative to the mean $l$ and $b$ values across stars in the high-quality member sample. We use the mean distance of the entire population for this conversion to avoid contaminating the sky positions, which have negligible uncertainties, with the uncertain distances. This puts both the sky position and distance from Earth in units of pc, however an additional scaling factor is necessary due to the large distance uncertainties. Comparing the on-sky extents of 3.5 pc in the $l$ direction and 3.4 pc in the $b$ direction in the clump discussed above with the distance standard deviation of 17.8 pc, we divide the distance by 5 to match their scales. For converting between position and velocity, previous SPYGLASS publications used a factor of 6 pc km$^{-1}$ s. However, in Circinus the much higher transverse velocity uncertainties inflate the velocity dispersion, necessitating a smaller conversion factor. In our test clump discussed above, we found standard deviations of 0.85 km s$^{-1}$ and 0.75 km s$^{-1}$ in $l$ and $b$ transverse velocity, respectively, relative to 3.5 and 3.4 pc in $l$ and $b$ position. We therefore select a conversion factor of 4 pc km$^{-1}$s to match these scales. The final 5-D space used for clustering is therefore $(d_{l,0},d_{b,0},d/c_d,c_v*v_{T,l},c_v*v_{T,b})$, where $c_v=4$ pc km$^{-1}$s and $c_d = 5$. 

We cluster the resulting coordinate space using HDBSCAN's leaf mode, setting both {\tt min$\_$samples} and {\tt min$\_$cluster$\_$size} to 6, which set the smoothing scale and minimum cluster size, respectively. The value of 6 was previously chosen for \citetalias{Kerr22a}'s clustering of the Austral complex in 6-D space-velocity coordinates, and in that paper the relatively low value was chosen to counteract the strict restrictions that were required ensure radial velocities capable of accurate clustering of the populations in 6-D space-velocity coordinates. Due to the highly restrictive conditions for inclusion in the high-quality member sample, we view the restrictions applied to this dataset as analogous. 

HDBSCAN assumes the presence of a substantial background component, although the relative purity of our sample ensures that many subgroup members will be assigned to the background. We must therefore assign outlying members to their most likely parent subgroup to produce sufficiently complete datasets for computing population statistics and ages. To do this, we use the same parameter space as for clustering, and assign outlying members to the nearest population according to the clustering proximity metric $D_N$, or the distance to the Nth nearest member (e.g., see \citetalias{Kerr22b}). This metric emulates the metric used in HDBSCAN's clustering, where $N$ is analogous to the {\tt min$\_$samples} HDBSCAN input parameter. We assign stars to the parent subgroup with which a given star has the lowest $D_N$. The extents of the original HDBSCAN-defined groups are shown in Figure \ref{fig:outlyingassignment}, with outlying stars marked according to the group assigned to them through this procedure. While some outlying stars may belong in a background or even an unclustered member component, these populations are small and are unlikely to affect our measurements of subgroup properties.

The clustering result is shown in Figure \ref{fig:circinus}, dividing the population into 9 subgroups. None of the resulting groups closely follow dense gas, except in cases where embedded populations are known, indicating that the high-quality member sample successfully suppressed the field contamination in the region. The largest group we find is ASCC 79, to which we assign the catalog ID CIR-8 \citep{Kharchenko05}. Semi-contiguous with this cluster, we see four adjacent populations - CIR-5, CIR-6, CIR-7, and CIR-9. CIR-5 was previously identified as OC-0625 in \citet{Perren23}, while CIR-6 and 7 both overlap with UPK 610 from \citet{Sim19}. CIR-6 and CIR-7 are separated by $\sim$ 44 pc in distance on average and UPK 610 has a reported distance directly between those of the two overlaid subgroups, so UPK 610 appears to be a more broadly-defined population that contains both subgroups \citep{Sim19}. CIR-9 has no known literature equivalent. Outside of this central region, CIR-1 corresponds to UPK 607, and CIR-2 matches with UPK 604 \citep{Sim19}. Finally, CIR-3 and CIR-4 are also well known, and represent largely embedded populations inside the Cir-W and Cir-E clouds within the CMC, respectively \citep{Reipurth08-cir}. We compile the properties of these Circinus subgroups in Table \ref{tab:subgroups}. 

\begin{deluxetable*}{cccccccccccccccc}
\tablecolumns{16}
\tablewidth{0pt}
\tabletypesize{\scriptsize}
\tablecaption{Demographics and mean properties of the subgroups in Circinus, identified by their parent association and ID. We also provide properties for Cir-E subcomponents discussed in Section \ref{sec:cir4dyn}, as defined by the linear cut in Figure \ref{fig:cir4}}
\label{tab:subgroups}
\tablehead{
\colhead{ID} &
\colhead{Names} &
\colhead{N\tablenotemark{b}} &
\colhead{M\tablenotemark{b}} &
\colhead{RA} &
\colhead{Dec} &
\colhead{l} &
\colhead{b} &     
\colhead{d} &  
\colhead{$\overline{v_{T,l}}$} &  
\colhead{$\overline{v_{T,b}}$} &   
\colhead{$R_{hm}$} &  
\colhead{$\sigma_{1D}$} &  
\colhead{$\sigma_{vir}$} &
\colhead{Virial\tablenotemark{b}} &
\colhead{Age} \\
\colhead{} &
\colhead{} &
\colhead{} &
\colhead{(M$_{\odot}$)} &
\multicolumn{2}{c}{(deg)} &
\multicolumn{2}{c}{(deg)} &
\colhead{(pc)} &
\multicolumn{2}{c}{(km s$^{-1}$)} &
\colhead{(pc)} &
\multicolumn{2}{c}{(km s$^{-1}$)} &
\colhead{Ratio} &
\colhead{(Myr)}
}
\startdata
1 & UPK 607 & 133 &  80.8 & 225.5 & -57.7 & 319.7 &  0.9 & 903 & -18.4 & -6.4 & 4.2 & 0.56 & 0.13 &  3.1 $\pm$ 0.9 & 8.8 $\pm$ 0.5 \\
2 & UPK 604 & 257 & 107.3 & 224.8 & -59.8 & 318.4 & -0.9 & 790 & -22.6 & -4.2 & 8.3 & 0.53 & 0.11 &  3.6 $\pm$ 1.8 & 5.4 $\pm$ 0.6 \\
3 & Cir-W   & 164 &  62.5 & 225.5 & -63.0 & 317.1 & -3.8 & 777 & -20.6 & -5.1 & 9.4 & 1.18 & 0.08 & 11.0 $\pm$ 3.7 & 1.9 $\pm$ 0.8 \\
4 & Cir-E   & 437 & 158.2 & 228.7 & -62.6 & 318.6 & -4.2 & 809 & -16.1 & -6.5 & 9.7 & 1.37 & 0.12 &  8.2 $\pm$ 2.0 & 1.5 $\pm$ 0.3 \\
 & Cir-Ea & 235 & 73.5 & 228.7 & -63.0 & 318.4 & -4.5 & 810 & -16.8 & -5.8 & 7.7 & 0.96 & 0.09 & 7.5 $\pm$ 2.7 & 1.5 $\pm$ 0.4 \\
 & Cir-Eb & 201 & 84.6 & 228.7 & -62.2 & 318.8 & -3.8 & 808 & -19.2 & -6.8 & 8.2 & 1.64 & 0.09 & 12.3 $\pm$ 3.1 & 1.3 $\pm$ 0.7 \\
5 & OC-0625 & 299 & 126.8 & 231.7 & -61.1 & 320.6 & -3.7 & 821 & -19.8 & -7.5 & 6.9 & 0.72 & 0.13 &  4.1 $\pm$ 1.0 & 4.7 $\pm$ 0.6 \\
6 & UPK 610\tablenotemark{a} & 352 & 166.3 & 229.6 & -59.3 & 320.7 & -1.6 & 830 & -16.9 & -8.9 & 2.0 & 0.52 & 0.27 &  1.4 $\pm$ 0.5 & 4.1 $\pm$ 0.9 \\
7 & UPK 610\tablenotemark{a} & 191 &  63.8 & 229.7 & -59.3 & 320.8 & -1.7 & 874 & -15.7 & -3.0 & 1.9 & 0.66 & 0.17 &  2.7 $\pm$ 1.1 & 2.7 $\pm$ 1.2 \\
8 & ASCC 79 & 933 & 501.1 & 229.5 & -61.0 & 319.8 & -3.0 & 823 & -17.6 & -5.6 & 5.3 & 0.50 & 0.29 &  1.2 $\pm$ 0.3 & 5.5 $\pm$ 0.6 \\
9 & & 363 & 219.6 & 230.0 & -60.2 & 320.4 & -2.5 & 811 & -19.3 & -7.3 & 6.3 & 0.72 & 0.17 &  2.9 $\pm$ 0.7 & 5.9 $\pm$ 0.5 \\
\enddata
 \tablenotetext{a}{UPK 610 contains CIR-6 and CIR-7, which are along nearly the same sight line.}
 \tablenotetext{b}{Mass, number, and virial ratio may miss stars with high velocities or parallax uncertainties, so they are best treated as lower limits}
\vspace*{0.1in}
\
\end{deluxetable*}

\section{Masses and OB stars} \label{sec:obstars}

Masses are necessary for estimating the virial state of populations, as well as identifying massive O and B stars, which have an important role in sculpting the evolution of stellar populations. At formation, these massive stars produce strong feedback that clears gas from their immediate vicinity. This feedback can also compress nearby dense gas clouds, which can act as a star formation trigger \citep[e.g.,][]{Elmegreen77}. The role that these stars play in guiding the evolution of gas makes them essential to our understanding of star formation. 

We measure mass by comparing the photometry of each candidate Circinus member to a set of isomass tracks from the PARSEC v1.2S isochrone models \citep{PARSECChen15}, correcting for extinction using the \citet{Lallement19} maps. We use the same grid as in \citetalias{Kerr24}, which has mass sampling every 0.005 $M_{\odot}$ for $0.09 M_{\odot}<M< 1 M_{\odot}$, every 0.01 $M_{\odot}$ between $1.0 M_{\odot}<M< 2.0 M_{\odot}$, every 0.02 $M_{\odot}$ for $2 M_{\odot}<M< 4 M_{\odot}$, and every 0.05 $M_{\odot}$ for $4 M_{\odot}<M< 20 M_{\odot}$. We assign stellar masses according to the mass of the nearest isochrone track. 

As per \citet{Pecaut13}, main sequence O and B stars have masses M $> 2.7 $M$_{\odot}$, and we mark these stars in Figure \ref{fig:circinussky}, overlaid on the distribution of other members. We separately label particularly massive stars ($M>7 M_{\odot}$), as these stars are particularly important for driving gas evolution. We find that O and B stars are concentrated in and around ASCC 79 (CIR-8), with none connected to the CMC or UPK 604, and only one in UPK 607. The most massive stars in the sample are HD 135591 and HD 135160 at 13.45 $M_{\odot}$ and 11.5 $M_{\odot}$, respectively. These stars are located close to one another, with HD 135591 being found in ASCC 79, and HD 135160 in CIR-9. 

$\delta$ Cir is also a possible ASCC 79 member. It has velocities consistent with other Circinus members, but is excluded from our initial sample primarily due to its relatively close median geometric distance of 704 pc \citep{BailerJones21}. However, this distance is quite uncertain, and the 84th percentile of its distance posterior at 799 pc would put it close to the center of ASCC 79. With a mass of $20 \pm 0.5 M_{\odot}$ and spectral type O9IV, it is a subgiant with an expected main sequence lifespan similar to the age of its most probable parent subgroup \citep{Southworth22}. We therefore consider its membership in Circinus probable, given that there are very few other places where it could have formed in its short lifespan. With at least two companions, including an $11.41 \pm 0.24  M_{\odot}$ eclipsing companion and a $\sim 18.7 M_{\odot}$ more distant companion that may also be a binary, the $\delta$ Cir system has a total mass of over $50 M_{\odot}$ \citep{Mayer14, Southworth22}. As such, it may have had a key role in the initiation of later star formation bursts, and may even affect the gravitational binding of the cluster, as we later show that this system alone may account for up to 10\% of ASCC 79's mass. We mark the location of $\delta$ Cir in Figure \ref{fig:circinussky}. 

Most O and B stars in Circinus are located at the center of a gas void that appears in both the distribution of reddened subgiants in Figure \ref{fig:circinussky} and the \citet{Dobashi05} extinction maps, suggesting that most of the gas in the area is being actively cleared by these stars. After the two massive stars in the central region of Circinus and likely $\delta$ Cir, the next most massive star is the lone 7.7 $M_{\odot}$ star HD 132984, which is found in the outskirts of UPK 607. Its location suggests that it may have a role in sculpting the northern edge of the adjacent gas cloud, although more information on the precise distance to that cloud is necessary to properly assess that possibility. 

\section{Ages} \label{sec:ages}

A complete understanding of the star formation record in Circinus requires ages, which reveal the sequence of star formation events in the complex. We compute ages using the routine previously employed in \citetalias{Kerr24}, which fits the distribution of candidate members in Gaia $M_G$-($G_{BP}-G_{RP}$) color-magnitude space against a grid of PARSEC v1.2S isochrones. We perform this fit on the pre-main sequence over the range of colors $1.8 < G_{BP}-G_{RP} < 4$, with restrictions designed to limit the sample high-quality, reliable members without unresolved companions like those modeled by the isochrones. To remove low-quality measurements and likely contaminants, we only use stars in our high-quality member sample. This choice does exclude some probable members with quality photometry near the pre-main sequence turn-on, however the pre-main sequence in Circinus is high enough above the main sequence that this cut poses little risk of removing potential members in the section of the pre-main sequence where we perform the fit. We also remove stars with the Renormalized Unit Weight Error $RUWE>1.2$, which was shown by \citet{Bryson20} to remove most binary contaminants. We further restrict our samples to $RUWE < 1.1$ for any samples that would have at least 15 stars for fitting after applying that cut. This provides a maximally pure sample of single stars in all populations with a large enough stellar sample to enable reliable age fitting after imposing this restriction. 

\begin{figure}
\centering
\includegraphics[width=7cm]{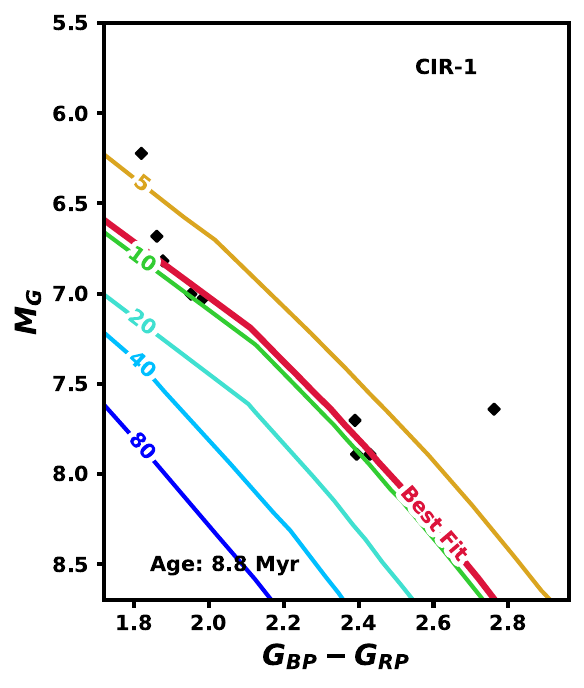}\hfill
\caption{Color-magnitude diagram showing the age fits for Circinus subgroups and CIR-4 subcomponents. Here we show CIR-1 (UPK 607) as an example, and the rest of the fits are available in the online-only version of this figure. The best fit isochrone is provided by the thick red curve, and isochrones of 5, 10, 20, 40, and 80 Myr are overlaid for reference (from top to bottom).}
\label{fig:agefits}
\end{figure}

Once the stellar sample is finalized, we apply a bootstrapping routine, selecting half of the stars at random (to a minimum sample size of 8), with the selection probability weighted by $P_{spatial}$. We then fit an isochrone to each bootstrapped sample. The final age solution is set by the 2-$\sigma$ clipped mean age across 10000 bootstrapped samples, while its uncertainty is provided by the standard deviation of the bootstrapped solutions.

\begin{figure*}
\centering
\includegraphics[width=17cm]{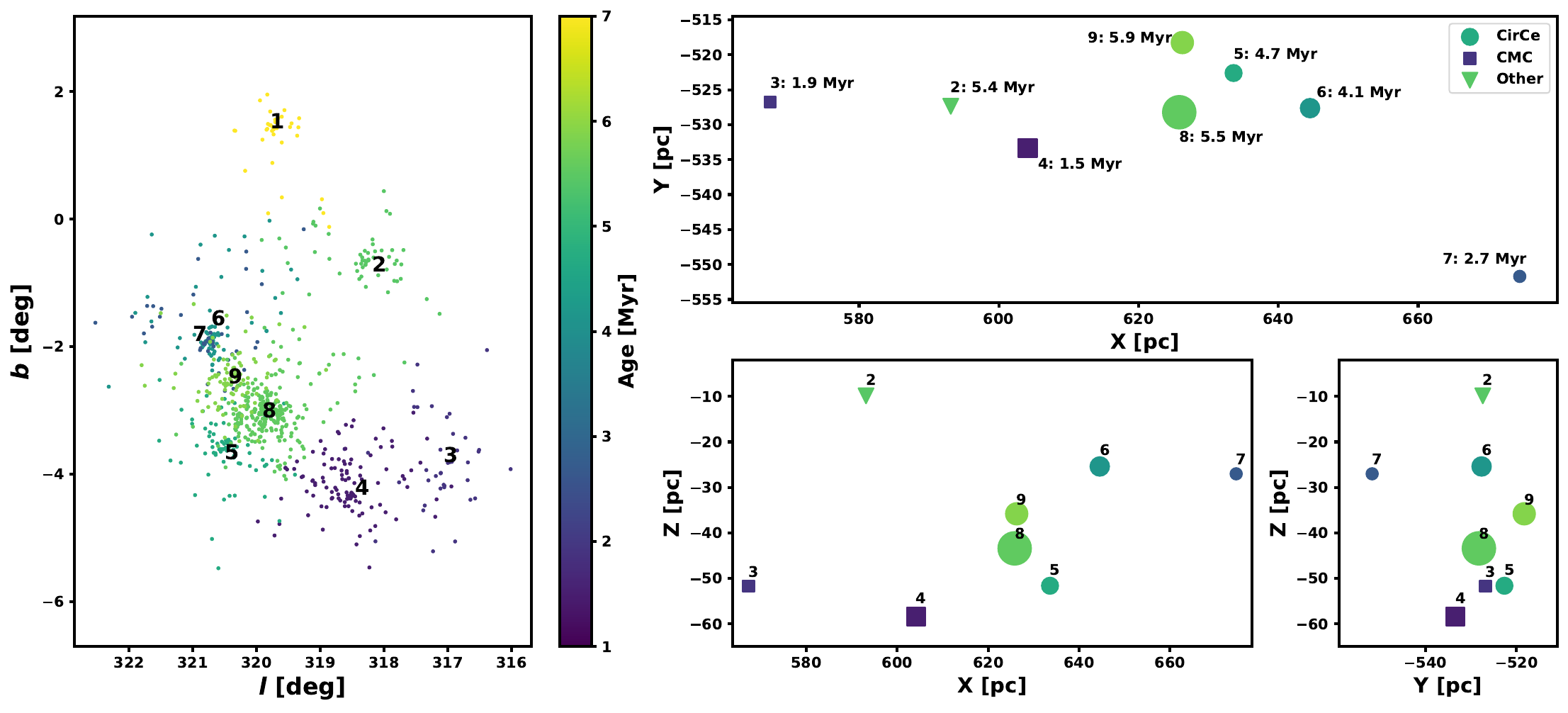}\hfill
\caption{In the left panel, a sky-coordinate map of the Circinus Complex, with subgroups colored by their ages. In the right three panels, a cross-section of the same distribution in galactic Cartesian coordinates, marking the mean position of each group rather than individual stars to suppress the visual impact of the high uncertainties in the distance axis. The markers are scaled with the mass of the subgroup (see Section \ref{sec:ages}). Circle markers are associated with the CirCe region (see Section \ref{sec:expansion}), squares with the CMC, and triangles with other populations. We exclude UPK 607 (CIR-1) from the right panels due to its outlying position. A 3D interactive version of the cross-section in the right panels is provided in the online-only version of this paper.}
\label{fig:circinusages}
\end{figure*}

We present the age solutions to each Circinus subgroup in the online-only version of Figure \ref{fig:agefits}, showing CIR-1 (UPK 607) as an example in-text. We also show the spatial distribution of those ages in Figure \ref{fig:circinusages}. Aside from the much older UPK 607 ($8.8 \pm 0.5$ Myr), ASCC 79 is the oldest population in the Circinus complex at $5.5 \pm 0.6$ Myr, alongside its similarly aged adjacent population, CIR-9. UPK 604 also has a similar age to ASCC 79, while CIR-3 and CIR-4 both show essentially newborn ages, consistent with their affiliation to the CMC. The other populations adjacent to ASCC 79, subgroups 5-7, are younger in proportion to their distance from that cluster, with the youngest age in the most distant contiguous subgroup, CIR-7, at $2.7 \pm 1.2$ Myr. This trend suggests that star formation in this central portion of Circinus started in ASCC 79 and propagated outwards. 

\section{Demographics} \label{sec:demographics}

To understand the scale of star formation in the Circinus complex relative to other young populations, we must gather basic information on the association's demographics. However, the Circinus region is sufficiently contaminated by field stars that our coverage of the population is likely to lack resolution at the lowest densities. Furthermore, stars with outlying positions caused by high distance uncertainties are likely to be excluded from our $P_{spatial}$ membership selection, and while high uncertainties are mainly an issue for dim stars, they can also be elevated for embedded or disk-bearing sources, which are common in Circinus. However, despite these limitations, we can use the populations that we have connected to the association to estimate scale of the association, with the caveat that it may underestimate the population's true scale. 

\subsection{Binaries} \label{sec:binaries}

Unresolved binary companions are a substantial reservoir of hidden stellar mass, and they must therefore be included to accurately measure the total demographics of an association. The most readily available test for unresolved multiplicity, \textit{Gaia}'s RUWE, can only identify binaries that are sufficiently separated that the point spread function in \textit{Gaia} differs from a single star. Especially at the distance of Circinus where the angular size of star systems is small, close binaries will often have a negligible effect on RUWE, making the population of unresolved binaries difficult to measure directly. Furthermore, many low-mass stars are completely invisible in \textit{Gaia}, making their binarity impossible to investigate. However, binarity statistics in our galaxy have been studied extensively, and those statistics can provide the mass and population in binaries \citep[e.g.,][]{Raghavan10, Duchene13, Sullivan21}. 

Before applying binary statistics, we must first produce a representative sample of system primaries, as the visible members of Circinus are heavily biased toward the massive stars that are visible through \textit{Gaia}. We do this by sampling $10^4$ stars at random from the \citet{Chabrier05} individual object IMF used in Section \ref{sec:lms}. Their binary statistics are computed following the choices of \citetalias{Kerr24}. Most companion fractions are gathered from \citet{Sullivan21}, with the exception of the high-mass bin centered on $M \simeq 3.25 M_{\odot}$, which we take from \citet{Duchene13} due to its absence in \citet{Sullivan21}. The mean mass ratios are defined by averaging the mass ratio distributions provided by the power laws in \citet{Sullivan21}. For both the binarity fractions and the mean mass ratios, the value for any given star is computed by interpolating between the values at the center of each mass bin from \citet{Sullivan21} and \citet{Duchene13}. 
The expected number of companion stars for a primary of a given mass is then provided by the companion fraction, while the missing mass in companions is provided by the companion fraction, multiplied by the mean mass ratio. The corresponding corrective factors are then the sum over mass and number for the input primaries and expected secondaries divided by the input primaries, which result in corrective factors of 1.41 in number of stars and 1.26 in mass. 

\subsection{Low-Mass Stars} \label{sec:lms}

\begin{figure}
\centering
\includegraphics[width=7.5cm]{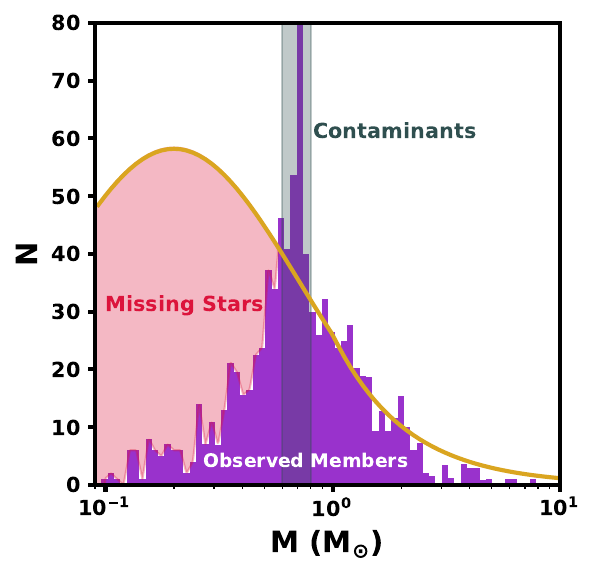}\hfill
\caption{Histogram showing the distribution of masses among observed stars in Circinus, shown in purple, plotted alongside the \citet{Chabrier05} individual object IMF, scaled to the histogram across the mass range $M > 0.8 M_{\odot}$, which is displayed as a yellow curve. The area to the left of the last crossover point between the histogram and scaled IMF is the volume occupied by missing stars, and in Section \ref{sec:lms} we use this area to compute corrective factors of 1.36 for mass and 2.28 for number. The gray shaded area marks a contaminant feature, which appears to be produced by a combination of model issues and field contaminants.}
\label{fig:IMFcorr}
\end{figure}

We estimate the number of stars excluded from our \textit{Gaia} data in Circinus by assuming that the members follow a standard Initial Mass Function (IMF), and scale that IMF to the distribution of high-mass members, which should be more complete. We present a histogram showing the distribution of masses in Circinus in Figure \ref{fig:IMFcorr}, where each star is weighted by its value of $P_{sp}$. Stars that lack a value of $P_{sp}$ are therefore excluded from this histogram. We also exclude resolved binary companions to avoid double-counting binary companions that we correct for in Section \ref{sec:binaries}. Our methods for identifying resolved binaries follow \citetalias{Kerr24}, and we summarize this methodology and show the sample of resolved binaries in Appendix \ref{app:bin}. We plot the resulting histogram against a version of the \citet{Chabrier05} individual object IMF, scaled to the histogram via least squares optimization for the mass range $M > 0.8 M_{\odot}$. Limiting this scale-matching fit to $M > 0.8 M_{\odot}$ avoids low-mass stars where the stellar distribution is less complete, along with the the large peak $M \simeq 0.7 M_{\odot}$. A similar peak has been seen in previous work on young stellar populations, and it appears to be driven by uncertainties in stellar model grids, in addition to the fact that isomass lines in the section of the CMD between the pre-main sequence and main sequence tend to be widely spaced, which can result in main sequence binaries and other photometric outliers piling up in this mass range, justifying its exclusion. A comparison of the histogram to the scaled IMF shows a substantial deficit of observed members among stars with $M < 0.6 M_{\odot}$ relative to the \citet{Chabrier05} IMF caused by \textit{Gaia}'s limited completeness among these dimmer sources.

A corrective factor can then be computed by comparing the size of the observed sample to that of the scaled IMF. We set the lowest-mass crossover point between the IMF and the histogram as the point where the mass deficit begins, so the missing population is the integral over the volume between the histogram and the IMF. The corresponding missing mass is the integral over that same area, with each bin element multiplied by its mass. We only consider stars with $M > 0.09 M_{\odot}$ for our stellar count, as this is the lower mass limit of the PARSEC isochrones, while all stellar masses are considered for our mass calculations. Corrective factors for the mass and population are provided by the sum over the observed and missing stars, divided by the sum over observed stars. The corrective factors for Circinus are 1.36 for mass and 2.28 for number. 

\subsection{Total Mass}

The total number of stars in a population is computed as the sum of $P_{sp}$ across all members, multiplied by the corrective factors for the number of low-mass stars and binaries. The corresponding mass is the sum over mass times $P_{sp}$, multiplied by the mass corrective factors for low-mass stars and binarity. These calculations produce a final mass for the association of 1484 $M_{\odot}$, spread across 3129 members. We select a systematic uncertainty of 20\% for these values, following typical discrepancies between different models of binarity and the IMF, in addition to the sensitivity of the IMF correction to biases given the limited mass coverage in Circinus. Circinus appears to have a scale similar to $\gamma$ Vel and Cinyras \citep{Pang21, Kerr24}, and we discuss its similarities to those populations in Section \ref{sec:comparisons}. However, we also acknowledge that the mass could be much higher, given the restrictive $P_{spatial}$ selection, and the presence of heavy extinction, which may result in the systematic underestimation of mass in gas-dense regions like the CMC where the \citet{Lallement19} extinction maps are less reliable. A more complete view of the Circinus complex will require improved modeling of the region's reddening to limit the contamination from subgiants, as well as radial velocities, which provide another velocity dimension in which to separate members from field stars. We summarize the populations by subgroup in Table \ref{tab:subgroups}.

\subsection{Boundedness}

To assess whether populations in Circinus are plausible open clusters, we compute a virial ratio for each subgroup. We follow the approach used in previous SPYGLASS papers and \citet{Kuhn19}, where bound clusters are defined as having $\sigma_{1D} < \sqrt{2}\sigma_{virial}$, and where $\sigma_{1D}$ is the square root of the average variance in multi-dimensional velocity-space. The virial velocity $\sigma_{virial}$ is defined as follows \citep{PortegiesZwart10}:

\begin{equation}
\sigma_{virial} = \left(\frac{GM}{r_{hm}\eta}\right)^{-\frac{1}{2}}
\end{equation}

\noindent where $M$ is the population mass, $r_{hm}$ is the half-mass radius, and $\eta$ defines the cluster's mass profile. Following \citetalias{Kerr24}, we take $\eta = 5$, which is a value typical of the broader profiles that are common in marginally bound young stellar populations. We compute $\sigma_{virial}$ and $\sigma_{1D}$ using sky-plane positions and velocities to compute $r_{hm}$ and $\sigma_{1D}$ following the methods employed in \citetalias{Kerr24}.

The resulting virial ratios are shown in Table \ref{tab:subgroups}, which are defined as 
$\sigma_{1D}/\sqrt{2}\sigma_{virial}$. A virial ratio less than 1 is bound. We find that while no Circinus subgroups are bound with confidence, both CIR-6 and ASCC 79 are within uncertainties of being bound. We therefore classify both as ``plausibly bound''. However, the probable underestimation of the mass in these populations due to the potential for uncorrected incompleteness at this distance, combined with the exclusion of high-uncertainty probable members like $\delta$ Cir, suggests that both populations are likely bound. 

\section{Dynamics} \label{sec:dynamics}

Associations are typically characterized by expansion signatures, as after the gas in a parent cloud disperses, the stars within become unbound, and drift apart according to their random motions inherited from the parent cloud. Expansion signatures can therefore mark stellar populations formed from a common site, which could have been anything from a single unbound group to a sparser network of small populations \citep[e.g.,][]{Wright20}. Due to a variety of factors such as progenitor cloud asymmetry, variable cloud dispersal timescales, and self-gravity, these expansion signatures are often anisotropic \citep{Wright24}. However, velocities that differ substantially from these expansion signatures may indicate populations that do not share an origin with the rest of the association. In this section, we present the dynamics of the Circinus Complex, search for expansion signatures, and discuss the nature of populations that follow different patterns. 

\subsection{Region-Wide Expansion Dynamics} \label{sec:expansion}

\begin{figure}
\centering
\includegraphics[width=8cm]{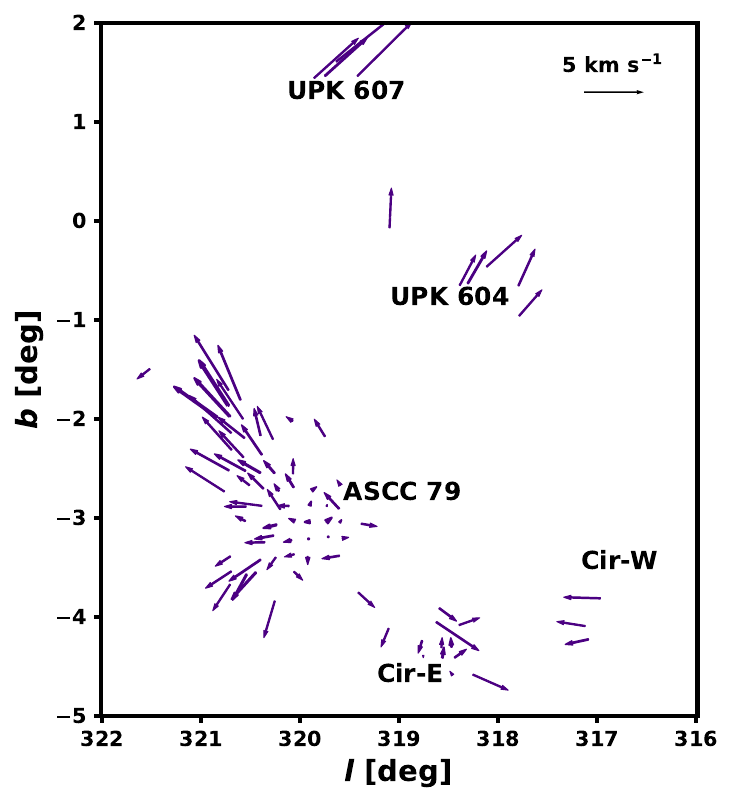}\hfill
\caption{Mean stellar transverse velocity directions and magnitudes across Circinus, corrected for projection effects ($v^{\prime}_{T}$). Values are calculated by dividing the area occupied by Circinus into 45 cells in $l$ and 50 cells in $b$, and computing average velocity in each cell. Velocities are only shown for grid cells with three or more stars, and the origin of each arrow is the average position of stars in that cell. The line weights are scaled by the number of stars in the cell.}
\label{fig:circinusvector}
\end{figure}

\begin{figure*}
\centering
\includegraphics[width=17cm]{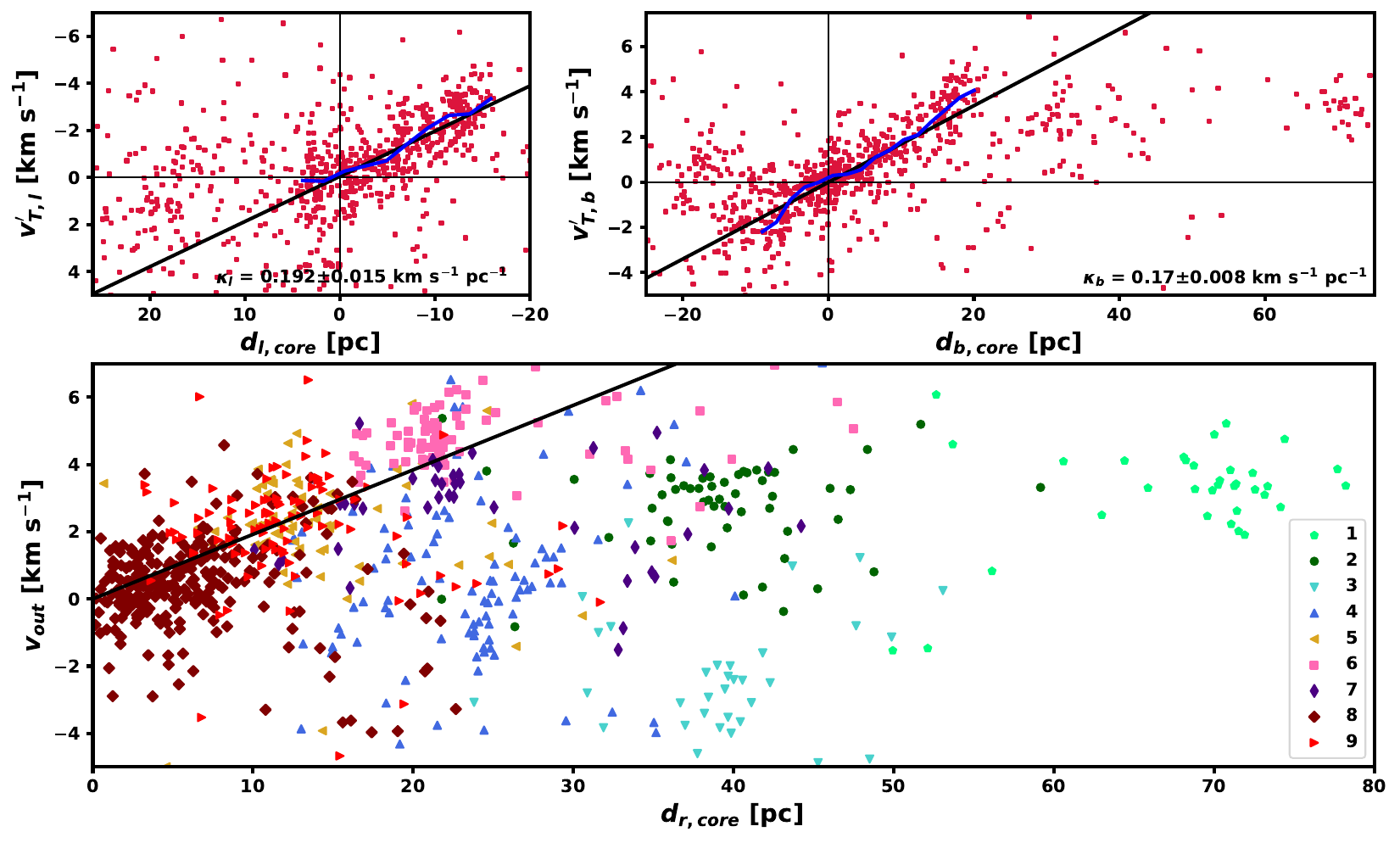}\hfill
\caption{Expansion signatures in Circinus. In the top row, transverse velocity in $l$ (left) and $b$ (right) corrected for virtual expansion, plotted against position along the $l$ and $b$ axes. Position and velocity are both given relative to the mean in the central cluster (ASCC 79). The axes for the $l$ plot are inverted to match the form of the $b$ plot, with CIR-6/7 to the right. We show fits to the position-velocity trends as black lines, with the best-fit expansion rate annotated in the lower-right. The blue curve shows the median velocity by bin in $l$ and $b$, which traces deviations from this fit. The bottom plot shows and velocities by Circinus subgroup, plotted using the markers from Figure \ref{fig:circinussky}. Velocities and positions are computed radially from the center of ASCC 79 in the plane of the sky, and the expansion fit in $l$ is shown for reference as a solid black line.}
\label{fig:circinusexpansion}
\end{figure*}

To provide an overview of the motions of stars in Circinus, we compute $v^{\prime}_T$, which is the transverse velocity adjusted for geometric effects associated with motion towards or away from the observer. When a population of stars is moving towards the observer, the angular size of the population increases as the population gets closer, resulting in a virtual expansion that must be removed to accurately measure the expansion of an association \citep{Kuhn19}. The corrected transverse velocity, $v^{\prime}_T$, accounts for these virtual effects, and is calculated using the following formulae:

\begin{align}
v^{\prime}_{T,l} &= v_{T,l} - \overline{v_{R}}*\cos b_c\sin(l_c-l)\\
v^{\prime}_{T,b} &= v_{T,b} - \overline{v_{R}}(\sin b_c \cos b - \sin b\cos b_c\cos(l_c -l))
\end{align}

We show the distribution of these corrected transverse velocities in Figure \ref{fig:circinusvector}, computed as an average across stars within each cell of an $l$ vs $b$ grid. The result shows a clear signature of expansion across groups CIR-5, CIR-6, CIR-7, ASCC 79, and CIR-9, which we hereafter refer to collectively as the Circinus Central (CirCe) Region. Stars have larger velocities further from the central cluster ASCC 79, and these more outlying locations are also where the youngest populations in the CirCe region reside. However, populations outside the CirCe region do not appear to have the same clear expansion pattern. UPK 604 has a nearly radial velocity vector relative to the ASCC 79, however it is also separated by over 40 pc without substantial populations connecting them. All other populations outside of the CirCe region, including those associated with the CMC, are generally not moving away from ASCC 79, indicating that they are all at least somewhat detached from the overall expansion signal. 

In the top row of Figure \ref{fig:circinusexpansion}, we show the distribution of $v^{\prime}_{T,l}$ and $v^{\prime}_{T,b}$ as a function of $l$ and $b$, respectively, providing two 1-D cross sections of the association's motion. Both axes show a distribution of velocities that consists of two components: a consistent trend of increasing relative velocities with increased distance from ASCC 79, and a scattering of velocities beyond that trend. We also see a tenuous flattening in the space-velocity slope near the core of ASCC 79, which also appears in the simulation we discuss in Section \ref{sec:starforge}. To quantify the expansion, we fit a line to the positions and velocities of stars in the CirCe region along the $l$ and $b$ axes, and include the results in Figure \ref{fig:circinusexpansion}. The best-fit slope has units of inverse time, so inverting the result provides the time since expansion started. We find that both axes show expansion timescales of $\tau \simeq 5$ Myr, albeit with minor asymmetry, with $\tau_{l} = 5.21 \pm 0.41$ Myr and $\tau_{b} = 5.88 \pm 0.28$ Myr. Both of these results match closely with the isochronal age in ASCC 79, suggesting that the CirCe region began its expansion around the time ASCC 79 formed, while younger populations like CIR-7 likely formed when the expansion of the region was already underway.

To produce a more detailed view of where this expansion signature applies, we plot radial distance and velocity in the plane of the sky relative to ASCC 79 in the bottom panel of Figure \ref{fig:circinusexpansion}, coloring each star by its parent subgroup. We also include the 5.21 Myr trend from the $l$ axis expansion fit for reference. Populations CIR-5 -- 9, which comprise the CirCe Region, all overlap with the expansion trend, and CIR-7, which is the most spatially and dynamically outlying of the set, has a 2-$\sigma$ clipped mean velocity offset only 1 km s$^{-1}$ below it. The remaining groups all reside much further below the expansion trend, indicating that their current-day velocities are not high enough to place the parent cloud in the vicinity of ASCC 79 in the $\sim$5 Myr since the initiation of star formation there. We discuss the potential origins of these populations in Section \ref{sec:nexpexpl}. 

\subsection{CIR-4} \label{sec:cir4dyn}

The Cir-E cloud has been shown to have at least two velocity components in $^{13}$CO emission, with the more extended northern component of the cloud having less negative values of $v_{LSR}$ compared to the center of the cloud \citep{Shimoikura11}. In our work, CIR-4, which is centered on the Cir-E cloud, has the highest velocity dispersion of any subgroup in the Circinus Complex, supporting the idea that substructure is present there. However, the strong and variable extinction in Cir-E raises the risk of contamination by reddened background subgiants, which may contribute to the large velocity dispersion in the region and overwhelm dynamical substructure. 

We therefore reassess the dynamical structure of the region by restricting the population to stars with $P_{Age<50 Myr} > 0.95$. This leaves only the stars with photometry almost entirely inconsistent with an origin on the main sequence, even when considering the full range of possible field metallicities, the presence of resolved main sequence binaries or triples, and typical reddening uncertainties. Only 44 of 114 high-quality CIR-4 members fail this highly restrictive cut, reinforcing the near-newborn nature of the population indicated by CIR-4's $1.5 \pm 0.3$ Myr bulk age solution.

\begin{figure*}
\centering
\includegraphics[width=18cm]{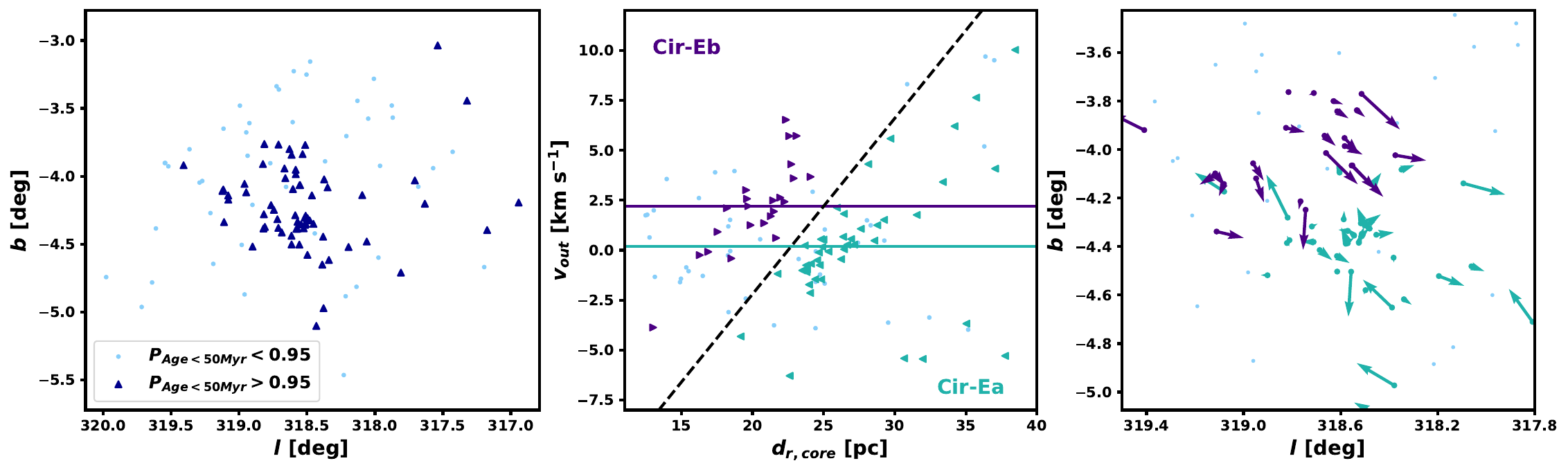}\hfill
\caption{Overview of CIR-4, where particularly strong contamination may mask substructure. The left panel plots stars that have $P_{Age<50 Myr} > 0.95$ (dark triangles) against those that do not (light dots). In the middle panel, we plot on-sky distance from ASCC 79 against the velocity along that separation vector, revealing two populations that separate cleanly after the removal of the lower-probability candidates. We split the sample into two components along the dashed black line. The sigma-clipped median velocity for each sub-component is plotted as a horizontal line, with the two sub-components separated by 2.0 km s$^{-1}$. In the right panel, we show the motions of stars in each sub-sample relative to ASCC 79, revealing that the component centered on the Cir-E cloud (Cir-Ea) has velocities very similar to ASCC 79, while the northern component (Cir-Eb) has velocities pointed away from ASCC 79 and towards Cir-E. }
\label{fig:cir4}
\end{figure*}

We provide an overview of the stellar populations in CIR-4 in Figure \ref{fig:cir4}. There we show that our cut disproportionately removes stars located far from the center of the region in spatial coordinates, and that the stars removed have very little concentration in velocity-space. These stars are therefore consistent with a background component, potentially mixed with stars that have evaporated from the older Circinus subgroups. We also show that the population separates cleanly along an axis in space-velocity coordinates, with one component consistent with the center of the Cir-E cloud, and the other offset to the galactic north, with some spatial overlap. The positions of these dynamical substructures match those of Cir-Ea and Cir-Eb from \citet{Shimoikura11}, respectively, so we label them accordingly. We include a quiver plot in Figure \ref{fig:cir4}, showing that Cir-Ea has a nearly identical velocity to ASCC 79 in the plane of the sky. Cir-Eb has a range of velocities pointed away from ASCC 79 and towards the core of the Cir-E molecular cloud, with velocities on the side furthest from ASCC 79 approaching those expected from the CirCe expansion signature. We compute age fits for each sub-component using the methods employed in Section \ref{sec:ages}, plotting the result in Figure \ref{fig:agefits} and listing the solutions in Table \ref{tab:subgroups}. Both components have ages near 1 Myr, reinforcing their affiliation with the surrounding gas.

The two components have sigma-clipped median velocities separated by 2.0 km s$^{-1}$, and unlike in the rest of the Circinus complex, these two populations are converging.  Both populations have protostars and gas associated with them \citep{Rector25}, so these populations may be tracing a collision between clouds in the gas phase, between the main component of Cir-E and a less dense or largely exhausted cloud that is either falling into Cir-E under gravitational forces or being accelerated towards Cir-E by feedback from ASCC 79 and surrounding O and B stars. We elaborate on this possibility in more detail in \citet{Rector25}, which discusses the gas morphology and protostar content of the CMC, and shows that a cloud-cloud collision may explain the considerably denser and less filamentary structure in Cir-E compared to Cir-W.

\section{Comparison to STARFORGE Simulations} \label{sec:starforge}

Most of the stellar mass in the Circinus Complex is found in the CirCe region, and like many other nearby populations \citep[e.g.,][]{Pang21, Kerr24, Posch24}, it hosts both a global expansion signature and an age pattern with young populations found further from the massive central ``progenitor cluster'' ASCC 79. Linear expansion signatures are expected for co-natal groups, as the distance a star travels from the formation origin in a given time directly corresponds to its initial velocity, however this does not necessarily apply in cases of multi-generational star formation, where sub-populations form independently. The origin of an expansion trend in the context of a population like Circinus with extensive substructure in spatial coordinates, velocity, and age, is therefore unclear. To help us interpret these signatures, we look for age gradients in realistic star formation simulations, and assess their applicability to CirCe region.

\subsection{Simulation Introduction and Selection}

Two key features of the CirCe region are its expansion signature discussed in Section \ref{sec:expansion}, which indicates a state of unbound dispersal, along with its age trend, with younger subgroups towards the population's fringes and an older, particularly massive cluster in the center. An analogous simulation must therefore reproduce these patterns, and host similar total masses, velocities, and age ranges. Current models of sequential star formation from \citet{Elmegreen77} and \citet{Posch23} rely on stellar feedback as a mechanism to propagate star formation beyond the first generation, so the inclusion of stellar feedback is likely necessary to produce the age pattern seen in the CirCe region. Any simulation used for comparison must also produce unbound populations, as bound populations do not exhibit global expansion, and therefore cannot replicate an expanding population like the CirCe region. Globally bound populations also tend to hierarchically assemble, mixing any non-coeval stellar populations that form and erasing age substructure \citep[e.g.,][]{CournoyerCloutier23, Polak24}. The STARFORGE framework, which we have full access to through the STARFORGE collaboration, provides the most comprehensive treatment of stellar feedback to date. STARFORGE also consistently produces unbound stellar complexes with expansion signatures \citep{Grudic22, Farias24}, making the framework well-suited for comparisons with an expanding structure like the CirCe region. STARFORGE simulations also run for up to 10 Myr with full hydrodynamics compared to the 2-3 Myr timescales that are common in studies of cluster assembly, ensuring that the cloud is given enough time to produce multiple stellar generations \citep[e.g.,][]{Guszejnov22a, CournoyerCloutier23}. 

STARFORGE is a numerical framework developed in Gizmo \citep{Hopkins15, Hopkins16} that models the formation of star clusters from the collapse and fragmentation of a giant molecular cloud until the dispersal of gas via stellar feedback and supernova explosions \citep{Guszejnov21, Grudic22}. STARFORGE simulations include all relevant feedback processes involved in star formation, such as protostellar outflows, stellar winds, radiation pressure, photoionization, and supernova explosions, which makes them the most realistic models currently available. For comparisons with Circinus, we used the STARFORGE fiducial simulations performed by \cite{Guszejnov22a} \citep[see also][]{Grudic21}, which model molecular cloud conditions typical of the solar neighborhood \citep{Lada20}. The fiducial simulation model a 20,000~M$_{\odot}$ initially uniform cloud of 10~pc radius. Clouds have solar metallicity and initially turbulent with a virial parameter of $\alpha_{\rm turb}=2$, partially supported by magnetic fields, and exposed to the external radiation field of the solar neighborhood \citep[see][for details]{Grudic22}. 

\begin{figure*}
\centering
\includegraphics[width=18cm]{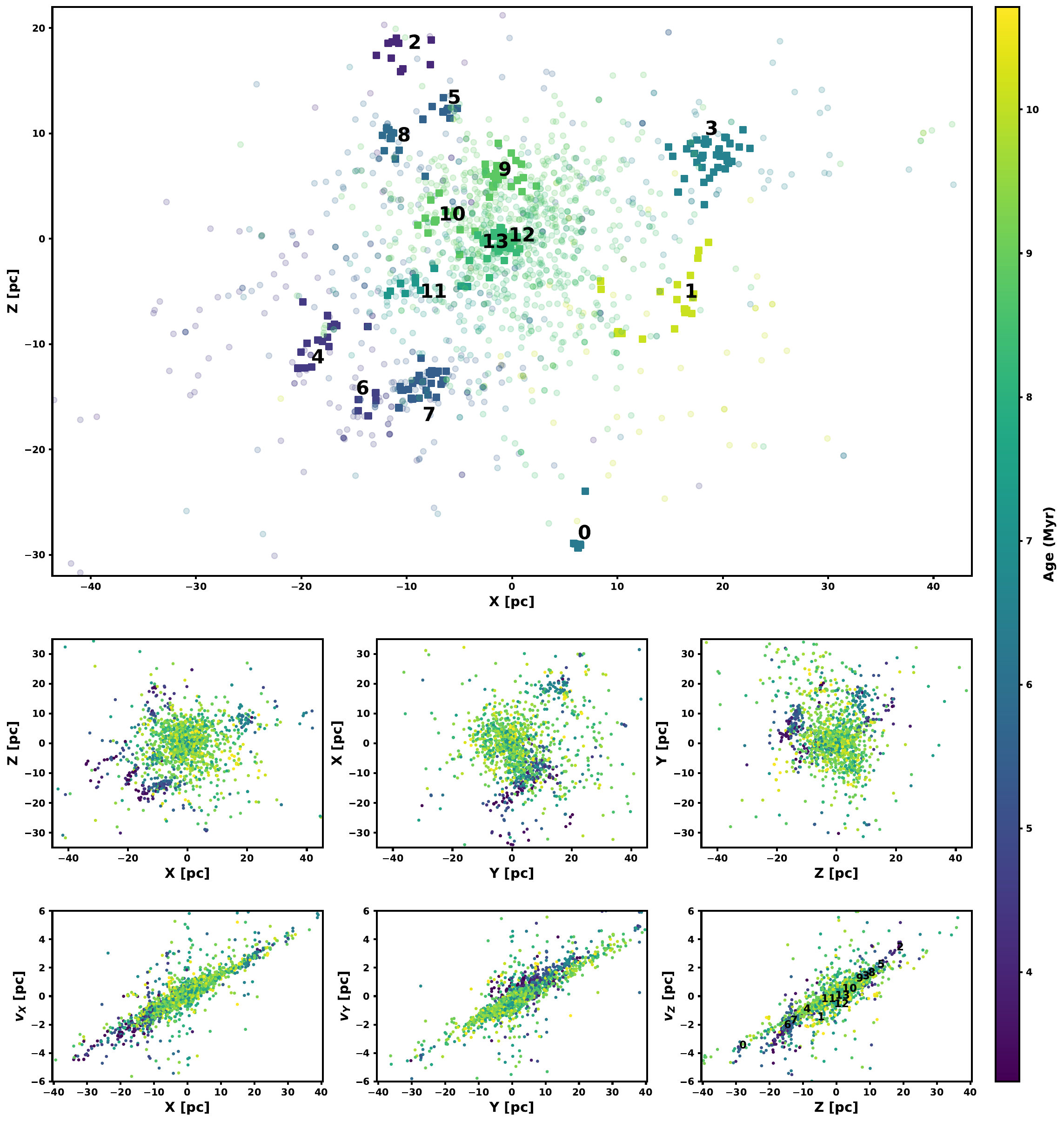}\hfill
\caption{Overview of F1 populations. In the top panel, F1 subgroups defined in $X$/$Z$ space, colored by their mean core age. HDBSCAN-defined members are marked by dark squares, while outlying members are marked with faded circles. In the bottom 6 panels we show a space-velocity cross-section of stars in the F1 simulation in space and velocity, colored by the simulated age of individual stars. Spatial dimensions are shown in the middle row, and space-velocity expansion trends like those in Fig. \ref{fig:circinusexpansion} are in the bottom row.  We annotate the mean position of subgroups in the $Z$/$v_Z$ panel to indicate the positions of the groups along the expansion sequence. As in Circinus, this simulation shows young populations near the edges, evidence for star formation sequences, and space-velocity trends consistent with expansion. }
\label{fig:F1xsec}
\end{figure*}

To locate a simulated age distribution reminiscent of the CirCe region, we looked at the stellar populations in the \citet{Guszejnov22a} STARFORGE simulations, which include three fiducial runs with different seed distributions, as well as simulations with variations on the properties of the fiducial run, such as a ten times higher initial density and an initially virialized cloud \citep{Grudic22,Guszejnov22b}. Through this search, the STARFORGE fiducial run from \cite{Guszejnov22a} with a random seed generator number of 1 (F1) emerged as a strong candidate for comparison. We provide a space-velocity cross-section of F1's simulated stars at a 12 Myr snapshot in the bottom two rows of Figure \ref{fig:F1xsec}, colored by their age. To better track the histories of specific populations, we perform HDBSCAN clustering on the simulated stars in F1. Since we have full 6-D phase-space data, we use the $(X,Y,Z,c*U,c*V,c*V)$ distance metric used for 6D clustering in \citetalias{Kerr22a}, with $c = 6$ pc km$^{-1}$ s. We then assign populations that are not initially included in an F1 simulation cluster (F1S) to the nearest population in that 6D parameter space. An overview of these subgroups is included in Figure \ref{fig:F1xsec}. 

Figure \ref{fig:F1xsec} shows that the the F1 simulation is dominated by an older central cluster surrounded by populations between 1 and 5 Myr younger, similar to what we see in Circinus. The velocities of subgroups in F1 span a contiguous range from 0 to 4 km s$^{-1}$ relative to the central cluster along a linear trend consistent with expansion. That velocity spread closely matches that of the CirCe region, and since the velocities and ages both correlate with distance from the central cluster, these similarities may indicate that similar forces are accelerating material in both scenarios, such as comparable levels of stellar feedback. The overall morphology and basic dynamical properties of F1 therefore meet our high-level requirements for a comparison with the CirCe region.

To further test the similarity of F1's expansion trend to the CirCe region, we fit a line to the space-velocity trends in $X$, $Y$, and $Z$ in F1 via least-squares optimization, and find expansion rates of $\kappa_X= 0.1317 \pm 0.0012$ km s$^{-1}$ pc$^{-1}$, $\kappa_Y= 0.1309 \pm 0.0023$ km s$^{-1}$ pc$^{-1}$, and $\kappa_Z= 0.1400 \pm 0.0030$ km s$^{-1}$ pc$^{-1}$. There is therefore some axis-dependent asymmetry in the expansion rates beyond $\kappa$ uncertainties, continuing a trend of asymmetric expansion in young associations that has become increasingly clear in recent studies \citep{Wright23}. The standard deviation of the two $\kappa$ measurements in the CirCe region is more than double that of F1, although due to the large uncertainties in $\kappa$ in the CirCe region and the domination of the region's asymmetry by CIR-6 and CIR-7, we conclude that the asymmetry of F1 is not inconsistent with it. The average of the three expansion rates in F1 would imply a time since the onset of dispersal of 7.5 Myr, approximately 1.2 Myr younger than the 8.7 Myr mean age of the component populations of the central cluster (see Sec. \ref{sec:simages}). Considering model-based systematic uncertainties in the isochronal ages, this is consistent with the near-zero difference between the isochronal age of ASCC 79 and the expansion age of the CirCe region \citep[e.g.,][]{Herczeg15}. The 8.1 Myr mean age across all stars in F1 at the 12 Myr snapshot is even closer to the dynamical age, showing that both CirCe and F1 have offsets between star formation and expansion much closer to zero than to the $5.5 \pm 1.1$ Myr offset observed in \citet{MiretRoig24}, which was proposed as a consequence of gradual gas dispersal.


Populations F1S-9, 10, 12, and 13 appear to trace fine substructure within the central cluster, and the total mass of these populations is 956 $M_{\odot}$, or 54\% of the total mass of the simulation. This share is very similar to the relative mass of ASCC 79, which contains 46\% of the mass of the CirCe region, although the total mass in F1 is 1784 M$_{\odot}$, approximately 60\% greater than the mass of the CirCe region. However, the younger outlying populations are more widely distributed around the central cluster in F1, while the younger populations in the CirCe region are distributed primarily along two axes towards CIR-6/7 and CIR-5. This discrepancy is not surprising given the spherical symmetry of the STARFORGE simulations, which may differ from the typically asymmetric environments where star formation occurs. However, the more distributed placement of the younger populations in F1 would likely change its appearance if it was placed at the location of Circinus by reducing the resolvability of the younger structures. The comparison between F1 and the CirCe region is therefore imperfect, but the overall structure, star formation sequence, and dynamics match closely.

\subsection{Simulated Population Ages} \label{sec:simages}

\begin{figure}
\centering
\includegraphics[width=7.5cm]{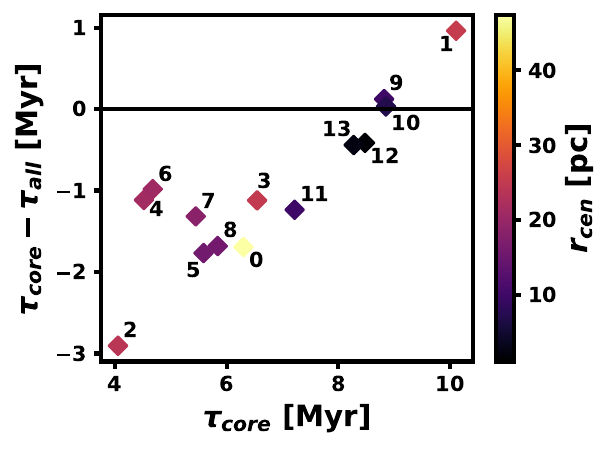}\hfill
\caption{The difference between F1 subgroup ages computed using core members ($\tau_{core}$) and all members including outliers ($\tau_{all}$), plotted against $\tau_{core}$. Populations are colored by their distance from the center of mass of the simulation and labeled. In young outlying populations, $\tau_{all}$ tends to underestimate the mean age among core members.}
\label{fig:dF1ages}
\end{figure}

All stars in F1 have ages associated with them, however for comparisons to Circinus, where we can only compute reliable ages for groups, group-wide ages are necessary. We therefore compute group ages in two different ways: one that takes the median age relative to the 12 Myr snapshot across all members of a simulated group, and another that includes only core stars defined as members by the initial clustering in that calculation. In Figure \ref{fig:dF1ages}, we show that these two definitions result in very different age results, with the solutions that use only the group core showing a wider range of ages relative to the solutions that include outlying members. This is caused by the fact that the velocities of stars that form in the central cluster and evaporate outwards are often near-indistinguishable from those of stars that formed during the expansion phase (see Sec. \ref{sec:starforgepatterns}), producing contaminants with ages that match the central cluster and can dominate the populations of younger subgroups when using the more inclusive membership definition. For later analyses and Figure \ref{fig:F1xsec}, we therefore use ages based on only core subgroup members, as they are more similar to the simulation-defined ages of the stars that the subgroups are centered on. 

Our isochronal age solutions in Circinus do address contamination effects by weighting the inclusion of a star in bootstrapping on its P$_{spatial}$, although more tenuous populations may be sufficiently contaminated by stars formed in the central cluster to substantially change the age solution. CIR-5, for example, appears to overlap with ASCC 79's halo, and has an age only 0.84 Myr younger than ASCC 79. This may be analogous to F1S-11, which has a median age 0.35 Myr younger than the central populations of F1S-12 and F1S-13 when all members are included, and an age solution 1.1 Myr younger when only core members are included. Our age differences between young subgroups and ASCC 79 may therefore underestimate the true age gap, which would leave more time for star formation triggering in a shell like in the F1 simulation. Furthermore, subgroup-level dynamical ages rely on pure samples, so the potential dominance of a central cluster may result in dynamical age solutions that measure the start of association-wide expansion, not local gas dispersal. 

\subsection{Star Formation Patterns in STARFORGE} \label{sec:starforgepatterns}

The formation of stars in F1 takes place in two stages: an initial phase of global collapse that forms most of the population, followed a expansion phase where smaller populations form in a residual gas shell \citep{Guszejnov22a}. Most populations in F1 that form while collapse is still ongoing merge into a central cluster through a process of hierarchical assembly \citep{VazquezSemadeni17}, but some small outlying populations drift independently, such as group F1S-1. These groups typically form first due to their shorter free-fall times compared to the more turbulent material near the cloud center \citep{VazquezSemadeni09}. F1S-1 has a substantial non-radial component to its velocity, although it still has a velocity consistent with expansion in two of three axes (see Fig. \ref{fig:F1xsec}). Due to its slightly older age compared to the central cluster, F1S-1 may be analogous to CIR-9, suggesting that CIR-9 is a leftover group from this initial assembly phase that never merged into ASCC 79. 

The collapse stage ends when newly formed O and B stars in the now fully-assembled central cluster begin to disperse the surrounding cloud, driving a new stage of global expansion. 
Near the central cluster, initially disorganized gas is swept up and compressed by a strong radiation front produced by the highly luminous young stars found there, triggering the formation of populations like F1S-3 and F1S-11. Further from the central cluster, organized structures often exist before the onset of gas dispersal, but they typically do not become star-forming until the arrival of the high-velocity gas front driven by the young O and B stars. Populations formed in the expansion phase have ages spanning from 4-7 Myr, with populations further from the central cluster being consistently younger. Along two different axes, this simulation hosts multi-generational star formation sequences, similar to what we see connecting ASCC 79 and CIR-7 in the CirCe region.

\begin{figure*}
\centering
\includegraphics[width=18.5cm]{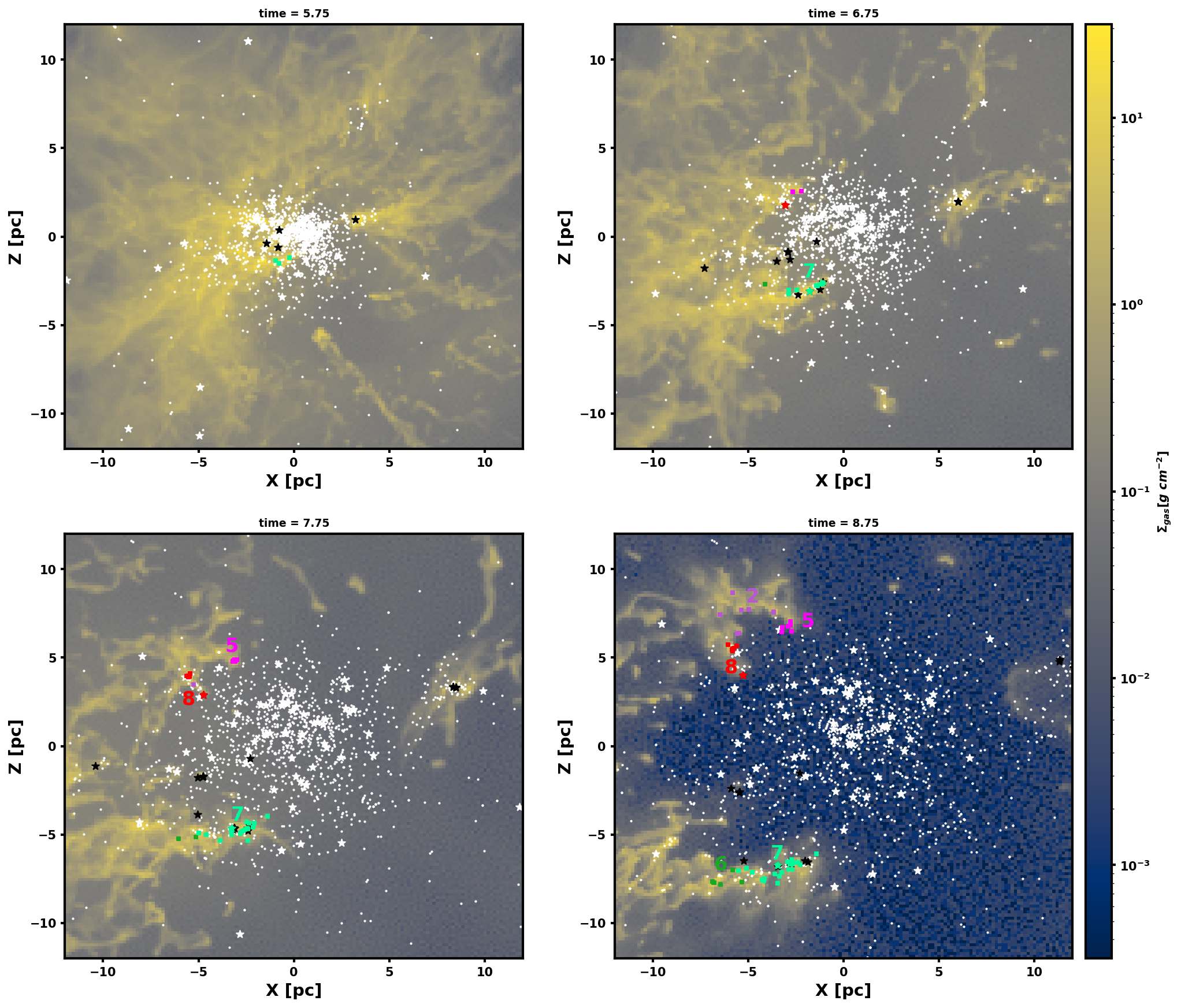}\hfill
\caption{The formation of stars in the STARFORGE F1 simulation, including both stars and gas. We highlight components of the two apparent star  formation sequences, marking HDBSCAN-assigned members formed later than 5 Myr of the five component groups: FIS-2 (purple), 5 (magenta), 8 (red), 6 (dark green) and 7 (light green). Other stars are marked in white. O and B stars ($M > 2.7M_{\odot}$) are given star markers, and these markers are black for O and B stars formed later than 5 Myr, flagging them as potential outlying members of these sequences. The colored background shows gas column density. F1S-5, 8, and 7 form before F1S-2 and 6, and none of them produce a substantial shock front that compresses the material that forms the next generation. The full animation is provided in the online-only version of this paper.}
\label{fig:F1feedbackgroups}
\end{figure*}

We highlight F1's star formation sequences in Figure \ref{fig:F1feedbackgroups}: Sequence 1, which consists of F1S-5, F1S-8, and F1S-2, and Sequence 2, which consists of F1S-6, and F1S-7. In Sequence 1, the core ages for F1S-5 and F1S-8 are 5.59 and 5.84 Myr respectively, compared to $\sim$8.7 Myr in F1's central cluster and 4.05 Myr in F1S-2. This produces an age sequence spanning nearly 5 Myr with three distinct generations along a largely linear axis. However, only one B star forms between F1S-5 and F1S-8, and the cloud hosting that B star remains largely intact until after F1S-2 forms. The lack of disruption to the parent cloud by this B star indicates that it is unlikely that it could have substantially changed the dynamical state of the gas that later formed F1S-2. However, during this same period, F1S-2's progenitor gas continues to be irradiated by the central cluster, and several O and B stars that migrated outwards from the central cluster closely interact with the material there. This suggests that this age gradient is driven entirely by the central cluster, with star formation initiating more slowly in clouds further from the central cluster. 

The dominance of the central cluster in driving star formation is supported by the structure of Sequence 2, where the component populations F1S-6 and F1S-7 have a similar age spread to the populations formed in Sequence 1. Unlike in Sequence 1, several O and B stars can be potentially connected to the older stellar generation of F1S-7. One of these stars manages to clear the adjacent section of the parent filament, however the section of the filament between that star and F1S-6 remains largely intact. It is not until the the material between that more distant section of the filament and the central cluster largely clears and the filament compresses that star formation initiates in F1S-6. 

The relative luminosities of stars in F1 support the dominance of particularly massive stars in the central cluster over less massive stars in later generations. In the F1 simulation, the most massive star is 57 $M_{\odot}$, while the most massive stars in the second-generation groups F1S-8 and F1S-7 are 5.3 and 3.6 $M_{\odot}$, respectively. Assuming a standard $L \propto M^{3.5}$ main sequence mass-luminosity relationship, the second-generation stars would need to be between 66 and 130 times closer to dominate over the massive cluster stars. This is far closer than the separations seen in Figure \ref{fig:F1feedbackgroups}, where F1S-2 and F1S-6 are typically no more than five times closer to the second generation populations than to the central cluster. Similarly, using the same luminosity relationship, the main sequence luminosity of the 20 $M_{\odot}$ $\delta$ Cir primary alone would be over 300 times that of the $\sim4 M_{\odot}$ stars at the center of CIR-6, so the second-generation stars would only dominate the incoming radiation for gas 20 times closer to those stars than to ASCC 79.  

\subsection{Triggered Formation in a Shell as an Explanation for Structure of CirCe region} \label{sec:shellform}

In this section, we have demonstrated that a pattern of global collapse and formation in a shell as shown in the STARFORGE F1 simulation is capable of producing an expansion signature similar to what we see in Circinus, with massive stars in and around ASCC 79 such as $\delta$ Cir as the most likely drivers of this pattern. In this scenario, only the old generation that built the central cluster exhibits true expansion. The velocities of later generations are produced by the bulk motions of the parent gas, which is accelerated away from the central cluster by stellar feedback. Differences in gas density and acceleration may explain several recent papers that show consistently anisotropic expansion signatures in substructured associations \citep[e.g.,][]{Wright18, Armstrong22, Armstrong24}. 


We find that the feedback that gives these younger populations their higher velocities can also produce an inside-out star formation pattern reminiscent of sequential star formation, with old populations mainly residing in a central cluster like ASCC 79, and multiple generations of progressively younger populations with higher bulk velocities forming further out. The F1 simulation shows that these multi-generational age gradients can be explained entirely through differences in exposure to stellar feedback from the central cluster. Star formation is triggered rapidly in structures located closer to the central cluster, where gas is subject to intense feedback from the O and B stars as soon as those stars form. More distant structures are subject to weaker feedback and experience shock fronts from the central cluster later, causing star formation to commence later. This central cluster-driven sequential star formation provides an alternative to the typical conception of sequential star formation, in which the stars triggering the star formation come from the most recent and therefore closest generation to the cloud \citep{Elmegreen77, Shu87}.

With star formation sequences spanning 5 Myr and velocities in younger generations of up to 4 km s$^{-1}$ relative to the central cluster, F1 has bulk properties that closely follow those of Circinus, as well as other known sites of sequential star formation like $\gamma$ Velorum and  Cinyras \citep{Pang21, Kerr24}. However, other associations like Sco-Cen show star formation sequences spanning 10 Myr or more, and the cluster chains in Sco-Cen tend to have a linear orientations \citep{Posch24}. The explanation of these structures in \citet{Posch24} is that these cluster chains are caused by the continual acceleration of a cloud by the aggregate of stellar feedback and supernova shocks from all of the previous generations. In a scenario in which an entire formation sequence is lined up radially away from the progenitor cluster, this explanation may be more coherent than ours, as in our scenario, the more distant clouds that should form later generations may be shielded from stellar feedback from the central cluster by gas closer to the progenitor cluster. However, the F1 simulation showed that star formation sequences can emerge in clouds at an angle to this radial orientation. Figure \ref{fig:circinusages} shows Circinus has an offset between the axis connecting the later generations (in this case CIR-6 and 7), and the location of the central cluster. Such an offset is also seen in Cinyras and $\gamma$ Vel, which are the only proposed sites of sequential star formation we are aware of with velocities and age spreads similar to the F1 simulation. As such, this central cluster-driven form of sequential star formation may be particularly important in these relatively short-lived star formation sequences, especially those with a massive progenitor cluster. New simulations will be necessary to confirm whether the mechanism we see in the F1 simulation can also produce associations like Cinyras, where the later generations have scales similar to that of its progenitor cluster RSG 5 \citep{Kerr24}. 

This feedback-driven view of sequential star formation provides a plausible explanation for the formation of Circinus and other similar populations. However, since we only found this formation pattern in one simulation, we are unable to determine exactly what conditions and processes permit the formation of age sequences. For example, the initiation of star formation in the expansion phase may require a minimum set of stellar feedback processes, and these requirements may vary based on the cloud structure. A specific sequence of massive star formation may also be necessary to drive gas compression without shocking the outer layers of gas. The emergence of these sequences may therefore depend on the star formation prescription, which differs depending on the framework \citep[see e.g.][]{Wall2020,CournoyerCloutier23}. As such, other simulations may show entirely different formation mechanisms for age gradients to what we see in F1. We will address these questions in a future study. 

\section{Discussion} \label{sec:discussion}

\subsection{Origin of Groups Outside Expansion Signature} \label{sec:nexpexpl}

The populations outside the main expansion sequence have velocities that are too low for the progenitor material to have been co-spatial with the Circinus core at formation. As such, they cannot be explained through the same combination of global collapse, expansion, and triggered star formation in a shell that we have proposed in the CirCe region. Here we discuss what relationship, if any, that these outlying groups have to the CirCe region. 

\subsubsection{UPK 607}

UPK 607 (CIR-1) has the weakest connection of any subgroup to the CirCe region, as its velocity vector does not align with the separation vector to ASCC 79 (see Fig. \ref{fig:circinusvector}), and it is located over 100 pc away from that central cluster (see Fig. \ref{fig:circinus}). There are also few young stellar populations between the CirCe region and the center of UPK 607, so the populations do not appear to have a structural link to one another. We therefore conclude that UPK 607 does not share a parent cloud with the rest of Circinus, although the similar ages and velocities suggest that these populations may be related in the context of larger galactic patterns. Including UPK 607 in studies of Circinus, if also considering other nearby young populations in that age range, may therefore help to trace larger-scale patterns that preceded the onset of star formation in the CirCe region. 

\subsubsection{UPK 604}

UPK 604 (CIR-2) is located near UPK 607 in the plane of the sky and shares similar velocities, but these populations are located on opposite ends of the Circinus Complex, separated by nearly 200 pc in distance (see Fig. \ref{fig:circinus}). This largely rules out a connection to UPK 607, but a connection to the CirCe region is plausible. Like populations in the expansion sequence, UPK 604 has a velocity vector pointing radially away from ASCC 79, however it is traveling more slowly than the expansion trend would predict at its position. A linear extrapolation would imply a position about 20 pc from ASCC 79 at formation. This is a larger separation than is seen between F1's central cluster and the populations formed in the shell, although the distance is small enough that UPK 604's parent gas cloud could have been accelerated by stellar feedback from ASCC 79. However, the best fit age for UPK 604 is largely consistent with that of ASCC 79, leaving little time for UPK 604's progenitor gas to compress and for its star formation rate to peak. An early onset of massive star formation in Circinus relative to the low-mass stars we use for computing ages may make this timeline plausible, however it is also possible that UPK 604 formed alongside the Circinus core and acquired its velocity prior to the onset of star formation in the Circinus core, similar to F1S-1. Acquiring a radial velocity here would confirm whether the 3D velocity is consistent with acceleration from the CirCe region, or whether its apparently radial motions relative to ASCC 79 are a chance alignment. 

\subsubsection{Circinus Molecular Cloud}

The CMC contains two actively star-forming regions that accompany substantial gas clouds (CIR-3/Cir-W and CIR-4/Cir-E). These clouds are vulnerable to acceleration due to feedback from the CirCe region, however the velocities here are slow relative to ASCC 79, and do not point radially away from the feedback sources there. As such, stellar feedback from the Circinus core does not appear to dominate here, suggesting that star formation in the CMC progresses independently of the CirCe region. However, the bimodal velocities in Cir-E may suggest that at least part of that cloud may be influenced by stellar feedback, or even expanding shell material accelerated by feedback from ASCC 79. We discuss the potential origin of this bimodality in \citet{Rector25}, where we combine the stellar dynamics outlined in Section \ref{sec:cir4dyn} with an analysis of the gas content and HH objects in the region.

\subsection{Age Gradients}

The CirCe region hosts an age gradient, with older populations like ASCC 79 in the center and younger populations further out, and with the youngest and most distant populations from ASCC 79 typically having the highest relative velocities. These age gradients differ fundamentally from those shown in the simulations of open cluster assembly in \citet{Farias19}, which display an age gradient with younger stars in the center. This different form of age gradient appears to be dynamically driven, as in a scenario with long-lived star formation in a central core, which would be the case if stellar feedback is not efficient enough at clearing the region's natal gas \citep{Watkins19, Dinnbier20}, the stars that form first have the most time to disperse \citep{Getman2014b}. The simulations from \citet{Farias19} reproduce previous observations of nearby open clusters like those from \citet{Getman18}, which investigated a sample of 19 morphologically similar star clusters that exhibit age gradients and found that younger stars are concentrated in the center and older stars are found further out.

The opposite age gradients, with the oldest populations in the center, have emerged in many recent observations of associations in addition to our results in Circinus \citep[e.g.,][]{Pang21, Kerr24}, but F1 is the first simulation we are aware of that reconstructs this pattern. Unlike the pattern seen \citet{Getman18}, where star formation appears to be driven by dynamical evolution outwards from a gradually-forming central cluster \citep{Farias19}, the evolution of F1 appears to present a scenario where the initial burst of star formation is short-lived, and where feedback-driven star formation occurs during the expansion phase of the region, when it is in a low-density state. In this scenario, most of the star formation takes place in a single collapse, without any infalling gas or structure that supports continuous star formation in the core after dispersal begins \citep{Grudic20, Farias24}. After that initial collapse, new populations form under the influence of feedback from the central cluster, producing several subgroups that do not infall, but rather drift apart as components of a substructured association. Additional comparisons with simulations will be necessary to investigate the full range of cloud masses, feedback conditions, and cloud dynamics that can produce each of these two age gradient scenarios.


\subsection{Future Evolution} \label{sec:comparisons}

Our ability to predict the future structure of the Circinus complex is limited by our near-complete lack of radial velocities in the region. However, we can estimate the future scale of the association using the present-day positions and velocities of members in the plane of the sky. Only CIR-6 and ASCC 79 are plausibly bound and in both cases this status is marginal, so gravitational attraction has little impact beyond the cluster cores. CIR-6, which is the CirCe region subgroup located farthest from the Circinus core alongside CIR-7, has a velocity relative to ASCC 79 of 4 km s$^{-1}$, implying that that current separation of $\sim$25 pc would increase to 65 pc after 10 Myr, 105 pc after 20 Myr, and 145 pc after 30 Myr. The 10 Myr time step would produce a structure that is approximately analogous to $\gamma$ Velorum, which has a total mass estimate of 1805 $M_{\odot}$, and has a roughly 85 pc separation between the 12.1 Myr old $\gamma$ Velorum cluster and the 7.5 Myr old Huluwa 5, which is the youngest and most distant subgroup. The age spread, mass, and overall scale is therefore quite similar to what we would expect to see in Circinus after 10 Myr of evolution, with $\gamma$ Velorum being perhaps slightly larger both in terms of mass and overall spatial scale. 

The scales and velocity in Sco-Cen may make components of it reasonably analogous to Circinus, however the age spread there appears to be much wider, with ages relative to the apparent initial generation spanning nearly 20 Myr. As discussed in Section \ref{sec:shellform}, this wider age spread may suggest that its formation was driven by a more gradual form of sequential star formation like the \citet{Posch23} model. For the 30 Myr age bracket, the populations of the Cep-Her complex may produce a strong comparison to Circinus. The Cinyras association, for instance, has a very similar spread of ages along its apparent star formation sequence from RSG-5 to CINR-6, although its velocities and therefore spatial scale, are smaller than what we would expect for Circinus after 30 Myr. 

The other major $\sim 30$ Myr association in Cep-Her, Orpheus, has a more similar velocity spread to Circinus, and as such, present day distances of subgroups to the most massive population, $\delta$ Lyr, match closely with the 145 pc scale expected in Circinus \citep{Kerr24}. However, Orpheus lacks clear age gradients, and also shows evidence of multiple star formation origins within the larger complex. Similar structure may soon emerge in Circinus, where UPK 607, UPK 604, and the CMC all appear to diverging largely independently of the CirCe region. While Circinus does not host populations in the CirCe region with a substantial velocity vector towards the other populations in Circinus, if it did, those populations would quickly overtake all three populations, since these populations are moving away from ASCC 79 at a rate less than the expansion rate. UPK 604, for instance, which has a velocity relative to ASCC 79 1 km s$^{-1}$ less than CIR-6. As a result, if CIR-6 was offset in the direction of UPK 604, it would overtake UPK 604 in approximately 20 Myr, at which point it would be difficult to separate from the rest of Circinus. If CIR-6 was offset from ASCC 79 in the direction of the CMC, it would overlap with the Cir-E core within only a few Myr. As such, with only minor differences in the orientation of the CirCe region relative to the surrounding clouds, Circinus could have a structure very similar to Orpheus, with multiple expansion origins that blend together, making age trends harder to distinguish. 

The gas content of the region may further complicate the future evolution of the complex. CIR-3 and CIR-4 lie in the CMC, where star formation is ongoing, and another substantial gas filament lies to the north of the CirCe region and south of UPK 607 \citep{Dobashi05}. CIR-7, which is the youngest population in the CirCe region, is located close to the interface between the dense stellar populations and the edge of that northern gas structure, suggesting that this cloud may be the current frontier of future star formation in the region. \citet{Dobashi05} shows that many dense cores are known in this northern cloud, and some Circinus candidate members overlap with this cloud in areas that do not clearly connect to any subgroup (see Fig. \ref{fig:circinussky}). This northern cloud could therefore be an excellent target for future virial analyses like those presented by \citet{Kirk17} and \citet{Kerr19} to assess whether star formation is imminent in this region, as well as protostar and HH object surveys, which may show that star formation has already begun. Identifying protostars embedded in these clouds would also improve our constraints on the distance to this cloud, better contextualizing its relationship to the stellar populations in the area. Future star formation events could nonetheless add substantial complexity to the appearance of the Circinus complex that could change what older populations serve as a close analog.

\section{Conclusion} \label{sec:conclusion}

We have conducted the first comprehensive demographic and dynamical survey of the Circinus Complex, an expansive site of recent and ongoing star formation associated with the Circinus Molecular Cloud and ASCC 79. Using \textit{Gaia} astrometry and photometry, we refined the membership of the complex previously reported by \citetalias{Kerr23}, identifying and age-dating substructures within the larger complex, and identifying dynamical patterns. The results of our survey work can be summarized as follows:

\begin{enumerate}
    \item The Circinus Complex has an estimated mass of $1484 \pm 297$ $M_{\odot}$ distributed across $3129 \pm 626$ members, making it comparable in scale to major nearby associations like Vela OB2 and Cinyras.
    \item Stars in the Circinus Complex are concentrated in the dynamically coherent $1085 \pm 217$ $M_{\odot}$ CirCe Region, which hosts the plausible open clusters ASCC 79 and CIR-6.
    \item The CirCe region hosts an age gradient, with younger ages found further from ASCC 79, and with ASCC 79, CIR-6, and CIR-7 comprising a three-generation star formation sequence with ages spanning $\sim 3$ Myr. It also hosts a velocity gradient consistent with a 5 Myr old expansion signature centered on ASCC 79.
    \item Outside the CirCe Region, UPK 607 appears dynamically and physically distinct from the rest of Circinus. UPK 604 and the CMC have velocities indicating they may have interacted with the CirCe region during their formation, however they are not subject to the global expansion signature seen in the CirCe region. 
\end{enumerate}

To investigate the origin of the age and velocity gradients in the CirCe region, we compared it to the STARFORGE F1 simulation, leading to the following conclusions that inform our interpretation of the morphology of Circinus and other similar populations: 

\begin{enumerate}[label={\Alph*}.]
    \item The star formation pattern in the CirCe region can be reproduced in its entirety by triggered star formation in a shell swept up by feedback from ASCC 79 and its associated O and B stars. Age gradients often associated with sequential star formation may therefore be a natural outgrowth of triggered star formation in the presence of gas overdensities with varied exposure to stellar feedback. 
    \item Young simulated groups formed during the expansion phase in STARFORGE are often contaminated with stars that evaporate from the central cluster, implying that the ages of outlying populations in associations like Circinus with a large central cluster may be overestimated.
\end{enumerate}

Due to its connections with star-forming gas, Circinus provides a well-populated environment for future studies of the interplay between massive young stars and gas while the star formation process is still ongoing. The insight provided by our comparisons to STARFORGE enforce the importance of simulations in the interpretation of star formation patterns across a wide variety of young associations. 

\
\

The authors thank the anonymous referee, whose comments greatly improved this publication. The Dunlap Institute is funded through an endowment established by the David Dunlap family and the University of Toronto. RMPK thanks Bruno Alessi for providing access to his internal catalog of literature associations, which was used for cross-matching populations identified in this paper.

\vspace{5mm}
\facilities{Gaia}

\software{astropy \citep{Astropy13},\citep{Astropy18}}

\appendix 

\section{Binaries} \label{app:bin}

Our detection of binaries in Circinus follows the approach in \citetalias{Kerr24}, where binaries are identified by searching a 10,000 au radius around each star at its \textit{Gaia} distance \citep{BailerJones21}. We identify stars within that radius as companions if they have a transverse velocity difference $\Delta v_{T} < 3$ km s$^{-1}$ and a distance difference $\frac{\Delta \pi}{\pi} < 0.2 $. These restrictions ensure that proposed companions have relative velocities within the range expected for binaries, and that their distances are not inconsistent with each other. We include stars that fail the $P_{spatial}$ cut for completeness.

\begin{deluxetable}{ccccccc} 
\tablecolumns{7}
\tablewidth{0pt}
\tabletypesize{\scriptsize}
\tablecaption{Catalog of binaries in Circinus, including basic stellar properties for members of potential binary systems. Stars with the same system ID are in the same system. $R$ is the distance from the primary, and the primary has $R = 0$. The full catalog is available in the online-only version of this paper.}
\label{tab:binaries}
\tablehead{
\colhead{Gaia DR3 ID} &
\colhead{Sys ID} &
\colhead{$R$} &
\colhead{RA} &
\colhead{Dec} &
\colhead{d}&
\colhead{m$_G$} \\
\colhead{} &
\colhead{} &
\colhead{AU} &
\colhead{(deg)} &
\colhead{(deg)} &
\colhead{(pc)} &
\colhead{}
}
\startdata
5799481287411833088 & 0 & 6880 & 227.8898 & -68.4827 & 790.6 & 17.21\\
5799481287411832192 & 0 & 0 & 227.8865 & -68.4848 & 876.8 & 12.09 \\
5800211981611187712 & 1 & 0 & 224.9700 & -68.0922 & 871.9 & 13.94  \\
5800211981600358656 & 1 & 879 & 224.9694 & -68.0921 & 898.5 & 14.06  \\
5800311006378292096 & 2 & 0 & 225.7941 & -67.8960 & 867.3 & 16.41  \\
5800311006378291584 & 2 & 3072 & 225.7942 & -67.8969 & 922.1 & 16.64 \\
5800323375872070016 & 3 & 3430 & 226.5844 & -67.7365 & 900.3 & 19.16  \\
5800323375872076928 & 3 & 0 & 226.5825 & -67.7357 & 1028.3 & 18.88  \\
5800354780687522560 & 4 & 0 & 226.4341 & -67.2139 & 826.0 & 14.48  \\
5800354780676056192 & 4 & 1153 & 226.4334 & -67.2135 & 741.0 & 16.25  \\
5800363164455408512 & 5 & 0 & 223.4426 & -68.3811 & 755.5 & 16.71  \\
5800363164429541120 & 5 & 4024 & 223.4452 & -68.3822 & 763.7 & 18.14 \\
\enddata
\vspace*{0.1in}
\end{deluxetable}

\bibliography{sample631}{}
\bibliographystyle{aasjournal}

\end{document}